\documentclass[preprint,
 aip,
 amsmath,amssymb,
]{revtex4-1}

\usepackage{graphicx}
\usepackage{dcolumn}
\usepackage{bm}

\usepackage[utf8]{inputenc}
\usepackage[T1]{fontenc}
\usepackage{mathptmx}
\usepackage{etoolbox}

 \usepackage{amsmath,amssymb,amsfonts}%
 \usepackage{mathrsfs}%
 \usepackage{xcolor}%

 \usepackage{xcolor}

 \usepackage{gensymb}

 \usepackage[normalem]{ulem}
 \usepackage{float}

\usepackage{lineno}

\makeatletter
\def\@email#1#2{%
 \endgroup
 \patchcmd{\titleblock@produce}
  {\frontmatter@RRAPformat}
  {\frontmatter@RRAPformat{\produce@RRAP{*#1\href{mailto:#2}{#2}}}\frontmatter@RRAPformat}
  {}{}
}%
\makeatother

\begin{document}

\preprint{}

\title[]{First observation of turbulence-like state in dense algal suspensions}
\author{Prince Vibek Baruah}

\author{Nadia Bihari Padhan}
\altaffiliation[]{Institute of Scientific Computing, TU Dresden, Dresden, 01069, Germany.}

 \author{Biswajit Maji}

 \author{Rahul Pandit}
 
 \author{Prerna Sharma}
\altaffiliation[]{Department of Bioengineering, Indian Institute of Science, Bangalore, C V Raman Road, Bengaluru, 560012, India.}

\affiliation{Department of Physics, Indian Institute of Science, Bangalore, C V Raman Road, Bengaluru, 560012, India.}
 \email{prerna@iisc.ac.in}

\date{\today}

\begin{abstract}
Active turbulence arises typically in systems ranging from microorganisms and biopolymers to synthetic colloids, where chaotic flows are closely associated with motile topological defects in collectively swarming suspensions. Here, we report the first experimental observation of turbulence-like dynamics in a fundamentally different class of systems: dense monolayers of motile unicellular alga \textit{Chlamydomonas reinhardtii} that exhibit neither orientational order nor topological defects. Nevertheless, the system displays rich spatiotemporal flow patterns with pronounced small-scale intermittency. We uncover strongly non-Gaussian velocity distribution, a feature distinct from both bacterial and classical fluid turbulence. Furthermore, we observe power-law regimes in the kinetic energy spectra, characterized by unique scaling exponents. Not only do our results provide compelling evidence for active spatiotemporal chaos in systems devoid of nematic or polar structures, but they also challenge current theoretical models. Our work opens new avenues for understanding emergent dynamics in active-matter systems and suggests intriguing biological implications, including enhanced mixing and transport in dense cell suspensions.
\end{abstract}

\maketitle

\section{Introduction}

Turbulence, which is ubiquitous from astrophysical to cellular scales~\cite{Frisch-CUP,biskamp2003magnetohydrodynamic,pandit2009statistical,boffetta2012two,pandit2017overview,benzi2023lectures,alert2020universal}, continues to provide new and exciting challenges 
for physicists, engineers, and mathematicians. Nonequilibrium turbulence-like states, which have been found over the last decade or so 
in dense bacterial suspensions~\cite{wensink2012meso,dunkel2013fluid,bratanov2015new,wang2017bactmulitfrac,alert2021,KiranPRF2023,padhan2025suppression}, epithelial-cell monolayers~\cite{Lin2021energetics,blanch2018turbulent,Doostmohammadi2015-re}, sperm suspensions~\cite{creppy2015turbulence}, and microtubule-motor mixtures~\cite{Sanchez2012-or,guillamat2017taming,martinez2021scaling,maryshev2019dry} provide important recent examples of new types of turbulence seen in biological systems. This has been christened \textit{active turbulence} because such suspensions are \textit{active fluids} in which energy is injected into the 
fluid not by an external force, as in conventional fluid turbulence, but by the conversion of chemical sources of energy to kinetic energy by the 
constituents in the suspension~\cite{alert2021}. So far, all the different active fluids, biological or synthetic, that show turbulence also have swarming motion of the constituents that require nematic or polar order parameters~\cite{pandit2025particles,giomi2015geometry,Yashunsky2024-hr}. Recent theoretical studies have shown that, surprisingly, even active-scalar fluids can also display turbulent states with rich phenomenology~\cite{tiribocchi2015active,nardini2017entropy,pandit2025particles,padhan2024novel,padhan2023activity,Keta2024emerging}.

Here, we present the first experimental study of spatiotemporal chaos and intermittency in non-swarming \textit{dense} algal suspensions. In particular, 
we carry out experiments on suspensions of two types of \textit{Chlamydomonas reinhardtii} (henceforth, \textit{C. reinhardtii}), the wild type CC1690 (WT)  and the mutant CC2377 (mbo2). We then characterise the statistical properties of this turbulence-like dynamics using measures that are employed to study conventional fluid turbulence. 

We obtain kinetic-energy and algal-concentration spectra, longitudinal-velocity structure functions~\cite{Frisch-CUP,pandit2009statistical,boffetta2012two,pandit2017overview,BuariaPRL2023}, probability distribution functions (PDFs) of velocity components, length-scale dependent longitudinal-velocity increments, and the Okubo-Weiss parameter $\Lambda$, which distinguishes between vortical and extensional regions in a flow. These spectra  display power-law scaling regions as a function of the wavenumber $k$; and they indicate that kinetic-energy and concentration fluctuations are spread over a wide range of spatial scales. However, the power-law exponents that characterise these scaling regions are distinctly different 
from their fluid-turbulence and bacterial-turbulence counterparts~\cite{Frisch-CUP,pandit2009statistical,boffetta2012two,pandit2017overview,alert2021}, as are the PDFs and structure functions mentioned above. Thus, our investigations uncover a new type
of multi-scale active chaotic flow, and whose special statistical properties we elucidate below. Earlier experimental and theoretical studies~\cite{leptos2009dynamics,thiffeault2015distribution,gollub2011enhancement} have investigated enhanced tracer diffusion in \textit{dilute} suspensions of swimming eukaryotic swimmers like \textit{C. reinhardtii}. However, this enhanced tracer diffusion is more akin to Lagrangian chaos or passive-scalar intermittency in simple flows~\cite{dombre1986chaotic,warhaft2000passive,falkovich2001particles} than to the fully developed spatiotemporal chaos we have uncovered here.\\
As we increase the concentration of \textit{C. reinhardtii} cells, the temporal evolution of our dense algal suspensions slows down and the intensity of turbulence-like dynamics in our systems decreases. Therefore, it behooves us to explore whether our algal systems cross over from a chaotic state to an \textit{active algal glass}. To examine this possibility for the cells in our suspensions, we calculate several quantities that are used to characterize slow dynamics in active glasses~\cite{sadhukhan2024perspective, Mandal2016-go, Henkes2011-hn}, including (a) the mean-square displacement (MSD), (b) the self-intermediate scattering function $F_s(k,t)$, (c) the overlap function $Q(t)$, and (d) the four-point correlation function $\chi_4(t)$, which characterizes the dynamic heterogeneity in a glassy system.
Our systems exhibit dynamic heterogeneity and distinct slowing down as $\bar{\rho}$ increases, but they do not display all the characteristic properties of an active glass.

\begin{figure*}
    \centering
    \includegraphics[width=1\linewidth]{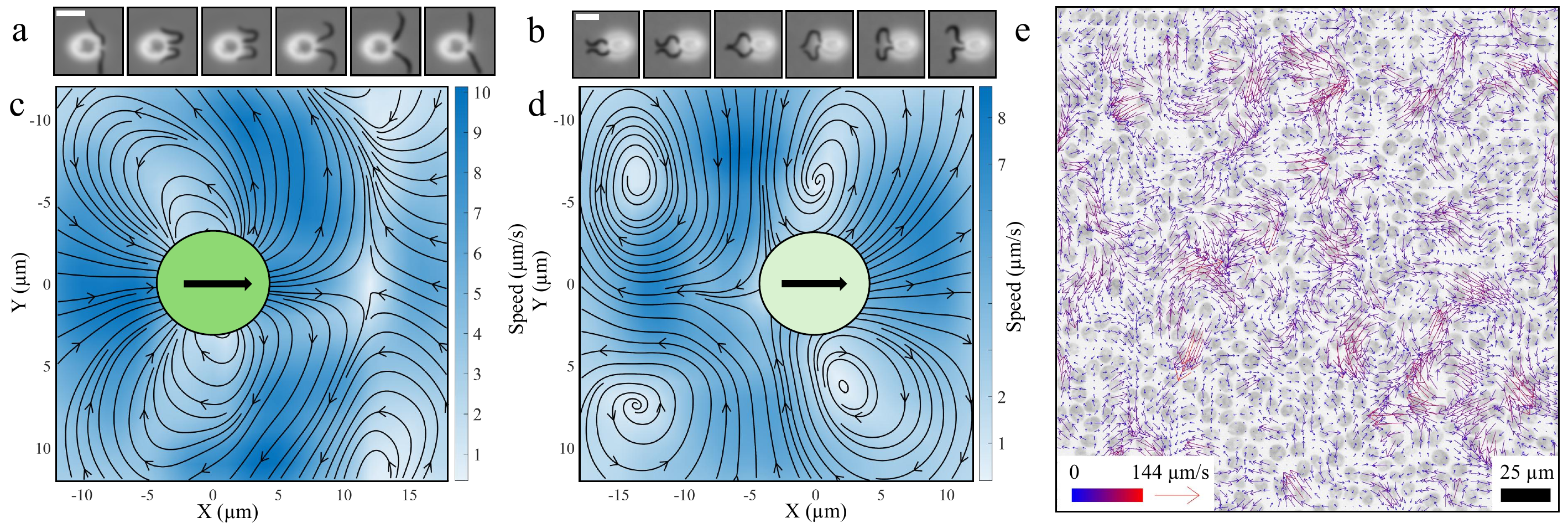}
    \caption{Time-lapse images of the motion of an isolated (a) wild-type (WT) and (b) mbo2  \textit{C. reinhardtii} cell [scale bar, $10$ microns]. The flagella of the cells in these images have been highlighted manually for clarity. 
    Experimentally measured beat-averaged flow-fields of isolated (c) WT and (d) mbo2 \textit{C. reinhardtii} cells; the arrow indicates the direction of motion. 
    (e) Interpolated velocity-vectors, overlaid on the image of a dense suspension of WT \textit{C. reinhardtii} cells. The supplementary material V7 shows its spatiotemporal evolution.}
    \label{fig:image_and_schematic}
\end{figure*}

\section{Statistical characterisation of dense algal suspensions}

Figure ~\ref{fig:image_and_schematic}  illustrates the two systems of wild type and flagellar mutant of \textit{C. reinhardtii}. Wild type \textit{C. reinhardtii} is a \textit{contractile} swimmer, on beat-averaged time scales and in the far-field limit [Figs.~\ref{fig:image_and_schematic}(a) and (c)]: it pulls the fluid from its front and back and pushes the fluid out from its sides. 
The two flagella of WT  \textit{C. reinhardtii} beat in a breast-stroke fashion as they propel the swimmer forward [Fig.~\ref{fig:image_and_schematic}(a)]. We compare and contrast the active spatiotemporal chaos of WT cells with that of the mutant mbo2 \textit {C. reinhardtii}, whose cells swim at a significantly lower speed ($v_m \simeq$ 50 $\mu$m/s) than those of WT \textit{C. reinhardtii} ($v_w \simeq$ 100 $\mu$m/s), and are effectively rear-propelled by the flagella. The beat-averaged flow field of  mbo2 cells resembles that of an \textit{extensile} swimmer in the far-field limit 
[Figs.~\ref{fig:image_and_schematic}(b), (d)]. We focus on collections of such swimmers, confined to a quasi-two-dimensional (2D) domain, wherein the cells swim in a wide chamber with a depth that is comparable to, but slightly larger than, the cell diameter [
Materials and Methods]. This allows us to visualize the spatiotemporal evolution of a monolayer of freely swimming \textit{C. reinhardtii} cells and we obtain the velocity vectors by tracking the individual cells between successive frames. We find that there is neither global nor local orientational or  nematic order present in these suspensions. The average polarization $P(t)=<\cos \theta >$, where $\theta$ is the angle between the instantaneous velocity vector of a cell relative to the $x$ axis, fluctuates around its mean value of zero. Therefore, there is no overall mean orientation in the system
(Appendix~\ref{appendix:orientational_order}). We also compute the probability distribution functions (PDFs)~\cite{Nishiguchi2015mesoscopic} of the orientational order parameter $| < e^{i\theta} >_{grid} |$ and the nematic order parameter $| < e^{2i\theta} >_{grid} |$
(Appendix~\ref{appendix:orientational_order}). The subscript $grid$ indicates an average over cells within a superimposed spatial grid; we consider several grid sizes. We do not see any peaks in these PDFs ~\cite{Nishiguchi2015mesoscopic}, so this confirms the absence of any local orientational or nematic order in the algal systems we consider. 

The swimmers collectively churn the fluid, which leads to an emergent chaotic flow, with more complex vortical fields than those generated by individual swimmers. We characterise this complexity systematically using statistical measures that are employed to analyse statistically homogeneous and isotropic turbulence in  fluids (see below). Figure~\ref{fig:image_and_schematic}(e) shows an optical image of the swimmers, at a representative time, overlaid with velocity vectors of the flow field, whose magnitude is given by the colour that goes from blue (low speed) to red (high speed). 
We note that the \textit{C. reinhardtii} cells have a low aspect ratio that is very close to 1 and is much smaller than in bacteria like \textit{B. subtilis}, whose aspect ratio varies from 4 to 8 ~\cite{ilkanaiv2017effect}. Consequently, there is no nematic phase in the absence of activity, and the \textit{C. reinhardtii} cells lack any intrinsic alignment because of their shape. This is also borne out by the rapid decay of the spatial velocity-velocity correlation function in our systems (Appendix~\ref{appendix:velocity_correlation}).

\begin{figure*}
    \centering
    \includegraphics[width=1\linewidth]{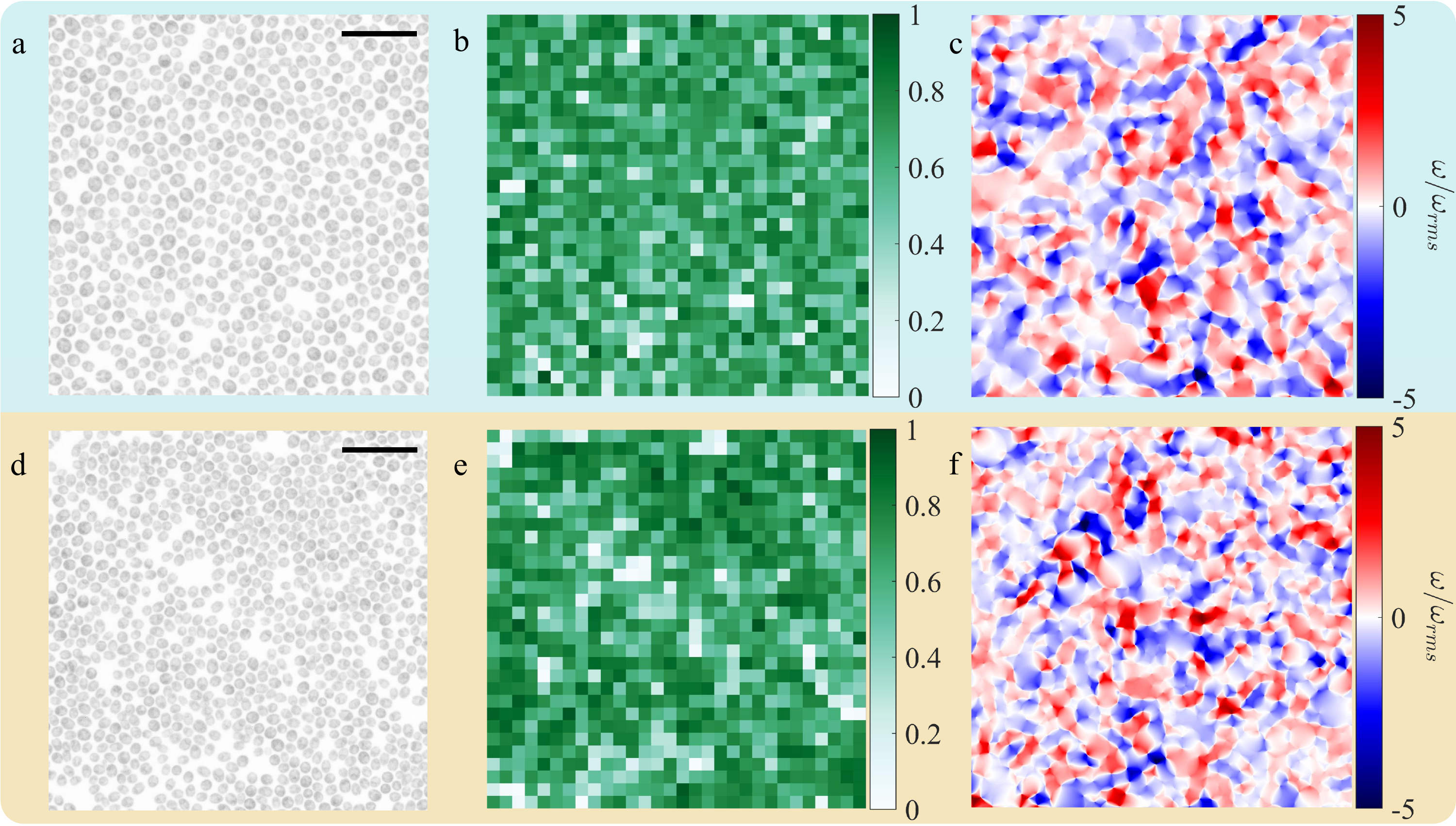}
    \caption{(a) Optical image of a suspension of WT \textit{C. reinhardtii} cells, with average density $\bar \rho = 0.63$ [scale bar, $50$ microns]; pseudocolor plots of (b) 
    the density and (c) the vorticity $\omega$ fields, for the image shown in (a). Panels (d), (e), and (f) are the mbo2 \textit{C. reinhardtii} 
    counterparts of (a), (b), and (c), respectively.}
    \label{fig:conc_vorticity}
\end{figure*}

\begin{figure*}
    \centering
    \includegraphics[width=1\linewidth]{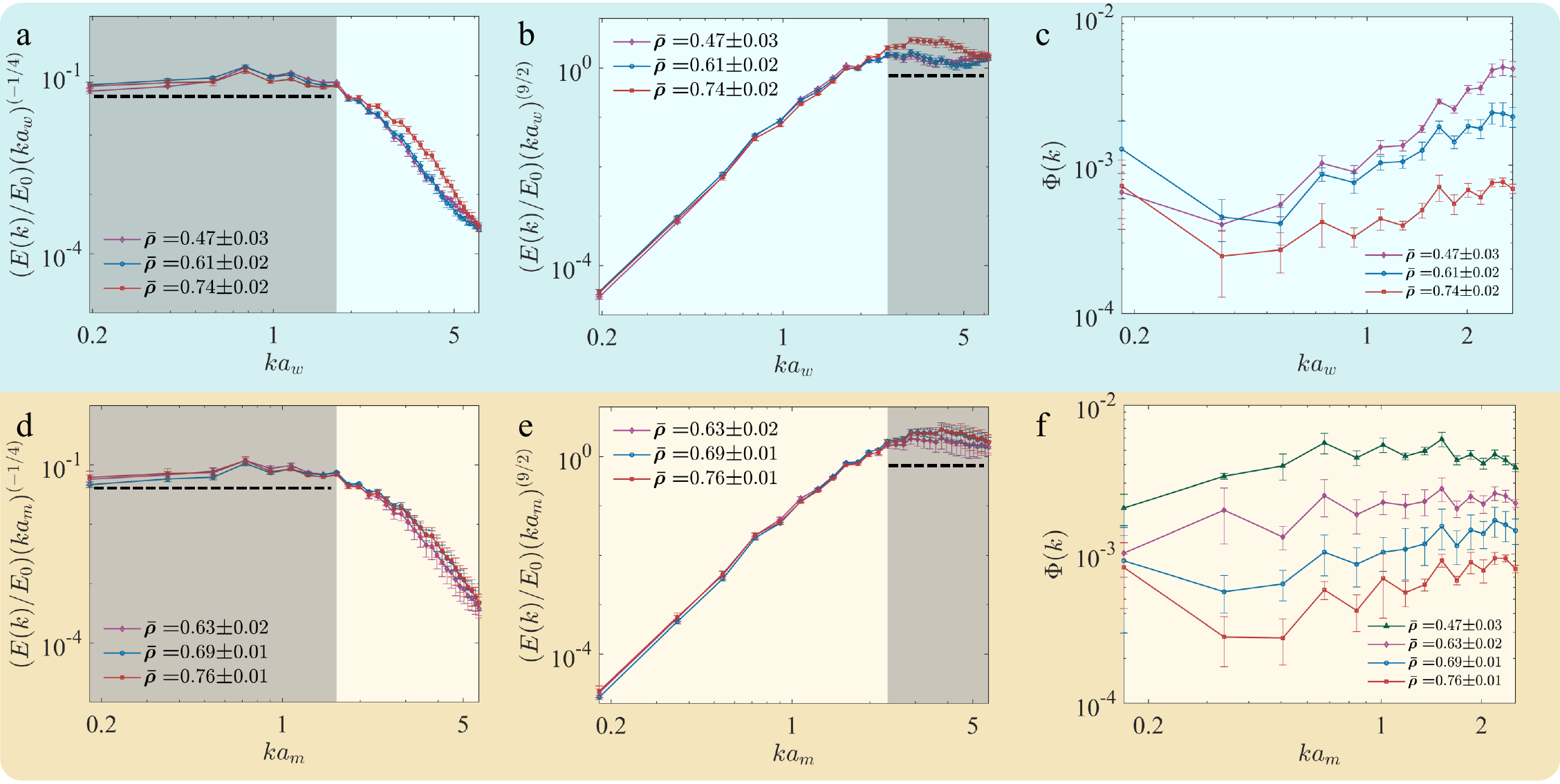}
    \caption{(a)-(b) Log-log plots of the energy spectra  $E(k)$, compensated by different powers of $k$ and plotted versus $ka_w$ for WT cells with $\bar \rho = 0.47, 0.61, 0.74$; (c) Log-log plots of the concentration spectra $\Phi(k)$, plotted versus $ka_w$. Panels (d), (e), and (f) are the mbo2 \textit{C. reinhardtii} counterparts of (a), (b), and (c), respectively. Dark-gray shading indicates scaling regions. Our data are averaged over different samples with similar values of $\bar\rho$ (within $\simeq 10\%$ of the mean); the error-bars denote one-standard-deviation ($\varsigma$), i.e., $\pm \varsigma (E(k)/E_0)(ka_w)^\frac{-1}{4})$, $\pm \varsigma ((E(k)/E_0)(ka_w)^\frac{9}{2})$ and $\pm \varsigma (\Phi(k)$. We define $E_0 = \sum E(k)$.}
    \label{fig:spectra}
\end{figure*}

We monitor the distributions of cells in dense suspensions of \textit{C. reinhardtii}, and show these in Fig.~\ref{fig:conc_vorticity} for WT cells (here and henceforth in a blue panel) and the mbo2 mutant (here and henceforth in a beige panel). Figure.~\ref{fig:conc_vorticity}(a) shows a microscopic snapshot of a  WT suspension with an average density $\bar\rho = 0.63$, at a representative time. At the same instant, we show a pseudocolor plot of the density field $\rho$, which lies between $0$ (no cell present) and $1$ (complete coverage with the cell bodies) [Fig.~\ref{fig:conc_vorticity}(b)] and the vorticity-field $\omega$ computed from the velocity field $\bm{u}$ [Fig.~\ref{fig:conc_vorticity}(c)]. Figures~\ref{fig:conc_vorticity} (d), (e), and (f)
are, respectively, the counterparts of Figs.~\ref{fig:conc_vorticity} (a), (b), and (c) for the mbo2 mutant. Supplementary material V1-V4 are raw experimental videos that show the spatiotemporal evolution of WT and mbo2 cell suspensions. Supplementary material V5 and V6 show the spatiotemporal evolution of the density and the vorticity fields (WT cells, $\bar\rho = 0.50$). These plots and the videos show that dense suspensions of both WT and mbo2 \textit{C. reinhardtii} cells show \textit{emergent nonequilibrium states} that appear to be turbulence-like in nature. We quantify the statistical properties of these states below.

The discrete velocity vectors from individual cells are interpolated over a finer equally-spaced grid [see Sec.~\ref{subsec:image_analysis}] and the resultant velocity field is Fourier transformed to compute the energy spectra.  Fig.~\ref{fig:spectra} shows log-log plots of the energy spectra $E(k)$ [Eqs.~\eqref{eq:Ekt} and \eqref{eq:Avspect} in Subsection~\ref{subsec:spectra}] and the concentration spectra $\Phi(k)$ [Eqs.~\eqref{eq:Skt},\eqref{eq:Avspect} in Subsection~\ref{subsec:spectra}] for the wild-type (blue panels) and mbo2 mutant (beige panels). We use $a_w$ ($a_m$), the mean wild-type (mbo2-mutant) cell size (Appendix~\ref{cell_size_section})
to non-dimensionalize  $k$. We also compensate the energy spectra with powers of $k$ to uncover different power-law regimes. The energy spectra scales as $E(k) \sim k^{1/4}$ in the small-$k$ limit and $E(k) \sim k^{-9/2}$ in the large-$k$ limit for all $\bar{\rho}$, and for both types of swimmers. But the density spectrum $\Phi(k)$ of WT and mbo2 cells differ from each other. This is because the spatial distribution of WT cells is different from that of the mbo2 mutant
(Appendix~\ref{appendix:triangulation} and Appendix~\ref{appendix:PDFrho}); and this distribution of mbo2 cells (at $\bar\rho = 0.50$), which are extensile swimmers, is somewhat reminiscent of phase separation ~\cite{padhan2024novel}.

There are significant differences between our spectra and their 2D-fluid-turbulence counterparts~\cite{pandit2017overview,boffetta2012two}. In brief, a conventional 2D fluid yields statistically homogeneous and isotropic turbulence, if it is forced sufficiently strongly at a length scale $l_f$; the resulting energy spectrum displays two spectral regimes -- the first with an \textit{inverse cascade} of energy and the second with a \textit{forward cascade} of enstrophy (i.e., the mean-square vorticity); in the inverse-cascade regime $E(k) \sim k^{-5/3}$, whereas, in the forward-cascade regime, $E(k) \sim k^{-\varpi}$, with $\varpi = 3$ in the absence of friction~\cite{pandit2009statistical,boffetta2012two}; the crossover between these two regimes occurs at a wavenumber $k_f \simeq 2\pi/l_f$. In the inverse-cascade regime, energy flows from the injection scale $l_f$ to larger length scales, in contrast to the forward cascade of energy in three-dimensional turbulence. The scale-dependent energy flux $\Pi(k)$ [Eq. ~\eqref{eq:Tketc}] is negative in the inverse-cascade regime, but positive in a forward-cascade regime~\cite{pandit2009statistical,boffetta2012two,pandit2017overview}.
  
Energy spectra for bacterial turbulence are also markedly different from those that we find for dense algal suspensions; e.g., the simple Toner-Tu-Swift-Hohenberg (TTSH) model, which has been used to model bacterial turbulence in \textit{Bacillus subtilis} and \textit{Escherichia coli}, yields a low-$k$ power-law regime in $E(k)$, but with an activity-dependent spectral exponent~\cite{bratanov2015new} that saturates eventually to a value $\simeq -3/2$ at large activity~\cite{mukherjee2023intermittency}.


We compute two natural characteristic length scales $L_E = 2\pi \frac{\sum E(k)/k}{\sum E(k)}$ and $L_\phi = 2\pi \frac{\sum \phi(k)}{\sum k \phi(k)}$ from the energy spectra and the density spectra
(Appendix~\ref{appendix:length_scales})~\cite{Perlekar2017two,Dhar1997some}. $L_E$ stays constant around $50 \mu m$ and $L_\phi$ stays constant around $25 \mu m$ for WT cells ($\bar\rho  = 0.50$). This suggests that the algal system is in a statistically steady state. Also, $L_\phi$ seems to increase monotonically with average density $\bar\rho$ for WT cells; but this behavior is not seen in mbo2 cells.

\begin{figure*}
    \centering
       \includegraphics[width=1\linewidth]{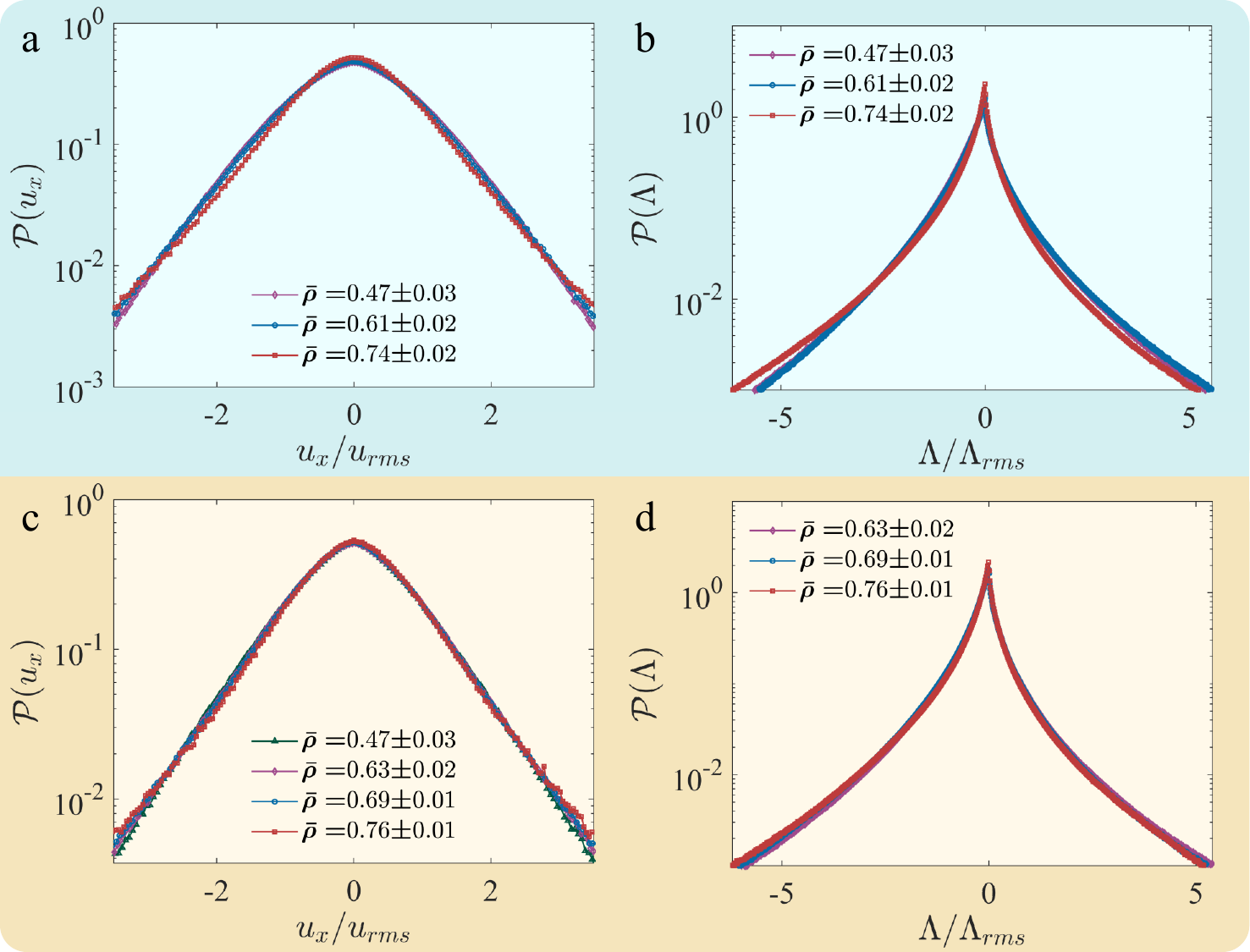}
    \caption{Semilog plots of different probability distribution functions (PDFs): PDFs of (a) the $x$-component of the velocity field $\bm{u}$ and (b) the Okubo-Weiss parameter $\Lambda$ [Eq.~\eqref{eq:Okubo}] for WT \textit{C. reinhardtii} cells and different mean densities $\bar{\rho}$. Panels (c) and (d) are the mbo2 \textit{C. reinhardtii} counterparts of (a) and (b), respectively.}
    \label{fig:PDFs}
\end{figure*}

Figure~\ref{fig:PDFs} [blue panel] presents PDFs ($\mathcal{P}$)   of (a) $\boldsymbol{u}_x$ and (b) the Okubo-Weiss parameter $\Lambda$  of WT cell suspensions  [Materials and Methods, Eq.~\eqref{eq:Okubo}~\cite{weiss1991dynamics,okubo1970horizontal,pandit2017overview}]. Their counterparts for the mbo2 mutant are given, respectively, in Figs.~\ref{fig:PDFs}(c) and (d) [beige panel]. The skewness for WT cells of $\bar\rho \sim 0.47, 0.61$, and $0.74$ are $0.29, 0.40$, and $-0.70$, respectively. For the mbo2 cells with $\bar\rho \sim 0.63, 0.69$, and $0.76$, the corresponding skewness values are $0.17, -0.32$, and $-0.70$, respectively. The PDFs for the wild-type system are qualitatively similar to their counterparts in the mbo2-mutant systems.

\begin{figure*}
    \centering
  \includegraphics[width=1\linewidth]{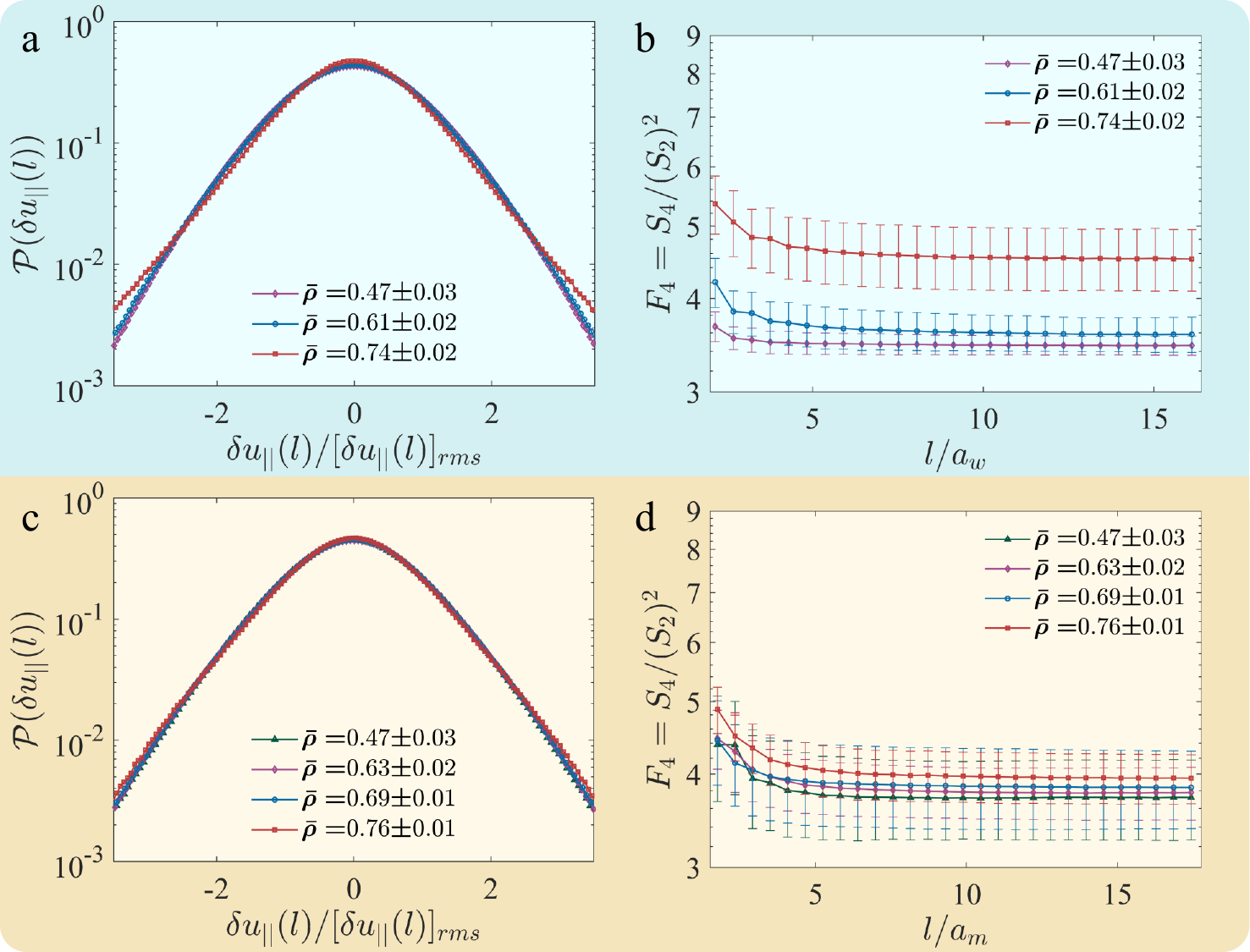}    
    \caption{ Semilog plots, for WT \textit{C. reinhardtii} cells, of (a) the PDFs of longitudinal-velocity increments of the velocity field $\bm{u}$, for different mean densities $\bar{\rho}$ and the separation $l/a_w = 5.37$. (b) the flatness $F_4$ [Eq.~\eqref{eq:flat}] versus $l/a_w$ for different values of $\bar{\rho}$. The plots in (c) and (d) are the mbo2-mutant counterparts of those in (a) and (b), respectively. Our data are averaged over different samples with similar values of $\bar\rho$ (within $\simeq 10\%$ of the mean); the error bars denote $\pm \varsigma(F_4)$.}
    \label{fig:flatness}
\end{figure*}

\begin{figure*}
    \centering
  \includegraphics[width=1\linewidth]{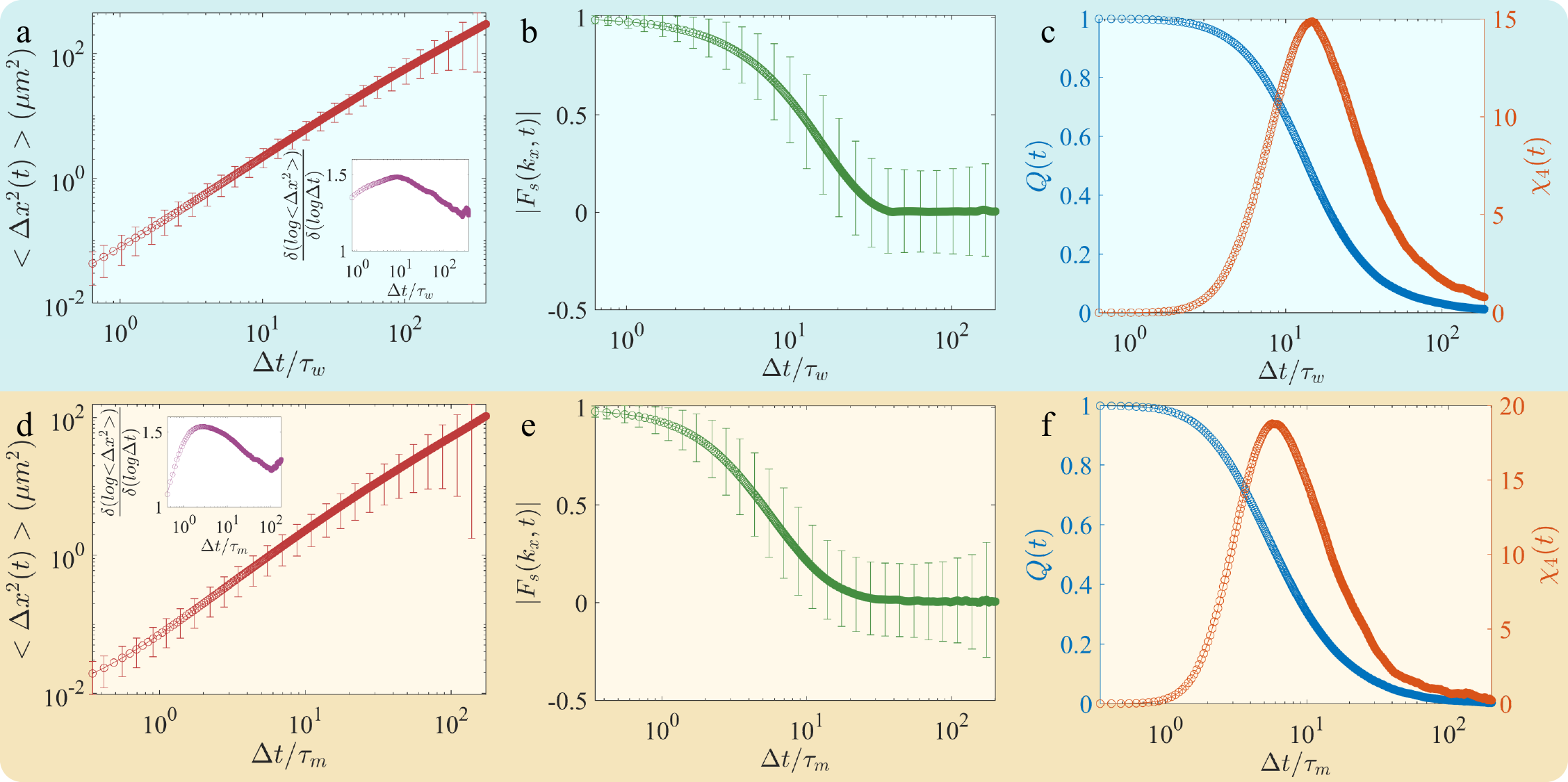}   
    \caption{Plots versus the scaled time $\Delta t/\tau_w$ of (a) the mean-square displacement (MSD) [the local slope $\frac{\delta(log \langle \Delta x^2 \rangle)}{\delta(log \Delta t )}$ is shown in the inset], (b) the modulus $|F_s(k_x,t)|$ of the self-intermediate-scattering function (for $k_xa_w =8.38$), and (c) the self-overlap function $Q(t)$ and the four-point correlation function $\chi_4(t)$ [Eqs.\eqref{eq:msd}, \eqref{eq:fskt}, \eqref{eq:qt}, \eqref{eq:chi4t}] for WT \textit{C. reinhardtii} cells with $\bar \rho = 0.75 \pm 0.02$; the reference time scale $\tau_w$ is defined in the text; (d), (e) and (f) are the mbo2-mutant counterparts of those in (a), (b) and (c), respectively. Our data are averaged over cell-trajectories of samples with similar values of $\bar\rho$ (within $\simeq 10\%$ of the mean); the error bars denote $\pm \varsigma(\Delta x^{2}(t))$ and $\pm |\varsigma(F_s(k_x,t))|$.}
    \label{fig:glass_main}
\end{figure*}

Note that the PDFs of $\boldsymbol{u}_x$ are \textit{markedly different} from those that are obtained for conventional fluid turbulence~\cite{pandit2017overview} and bacterial turbulence in dense suspensions of \textit{Bacillus subtilis}~\cite{dunkel2013fluid}, both of which are Gaussian. By contrast, the PDFs of $\boldsymbol{u}_x$, in Figs.~\ref{fig:PDFs}(a) and (c), are distinctly non-Gaussian (we fit them to compressed exponentials in Appendix~\ref{fit_section}).
The non-normalized flow speeds (Appendix~\ref{appendix:PDFvel})
decrease with an increase in cell concentration. Furthermore, $\mathcal{P}(\Lambda)$ of these dense suspensions of \textit{C. reinhardtii} cells is
significantly different from its counterparts in 2D fluid turbulence~\cite{Perlekar2009-kk} and bacterial turbulence~\cite{KiranPRF2023}.

Figure~\ref{fig:flatness}(a) [blue panel] displays  the PDF ($\mathcal{P}$) of the longitudinal velocity increments $\delta \boldsymbol{u}_{\parallel}$(l) of WT cell suspensions for different mean densities $\bar{\rho}$ and the separation $l/a_w = 5.37$ (Materials and Methods, Eq.~\eqref{eq:velinc}~\cite{pandit2017overview}). The separation length $l$ is the magnitude of the vector $\boldsymbol{l}$ that we use in the definition of the longitudinal velocity increment. With a fixed vector $\boldsymbol{l}$, we calculate the difference of the velocity $\delta \boldsymbol{u}_{\parallel}(l)$ along that direction [see Eq.~\eqref{eq:velinc}]. To generate the complete PDF, we take all possible directions for a fixed magnitude $l$. Figure~\ref{fig:flatness}(c) [beige panel] gives the corresponding PDF for the mbo2 mutant (here $l/a_m = 5.82$). The $l$-dependence of these velocity-increment PDFs are shown in Appendix~\ref{appendix:PDFincrement}.
This $l$-dependence is a fingerprint of intermittency, which we quantify by the flatness $F_4$ that is related to the fourth- and second-order velocity structure functions [Materials and Methods, Eq.~\eqref{eq:flat}].
We plot the flatness $F_4$, for both WT (versus $l/a_w$) and mbo2 (versus $l/a_m$) variants, in Figs.~\ref{fig:flatness} (b) and \ref{fig:flatness} (d), respectively. Both these plots show distinct deviations from the Gaussian value $F_4^G=3$; this deviation increases as $l/a_w$ (or $l/a_m$) decreases, a clear manifestation of small-scale intermittency~\footnote{Signatures of small-scale intermittency are often uncovered by such plots of flatness or the hyperflatness $F_6\equiv S_6/(S_2)^3$ in fluid and other forms of turbulence~\cite{Frisch-CUP,perlekar2006manifestations,sahoo2011systematics,pandit2017overview}}.

We turn now to an investigation of possible signatures of an active glass in our WT and mbo2 systems at large values of $\bar{\rho}$. Figures~\ref{fig:glass_main} (a), (b), and (c) show, respectively, plots of the mean-square displacement (MSD), the modulus $|F_s(k_x,t)|$ of the self-intermediate-scattering function (for $k_xa_w =8.38$), the self-overlap function $Q(t)$, and the four-point correlation function $\chi_4(t)$ versus the scaled time $\Delta t/\tau_w$ [Eqs.~\eqref{eq:msd}, \eqref{eq:fskt}, \eqref{eq:qt}, and~\eqref{eq:chi4t}] for WT \textit{C. reinhardtii} cells with $\bar \rho = 0.75$; the reference time scale $\tau_w\equiv a_w/v_w$,
where $v_w \simeq 100\; \mu {\rm {m/s}}$ is the mean swimming speed of isolated WT cells; the local slope $\frac{\delta(log \langle \Delta x^2 \rangle)}{\delta(log \Delta t )}$ is shown in the inset of Fig.~\ref{fig:glass_main} (a).  Figures~\ref{fig:glass_main} (d), (e), and (f) are the mbo2 counterparts of Figs.~\ref{fig:glass_main} (a), (b), and (c); for mbo2, $k_xa_m = 15.47$ and the reference time scale $\tau_m\equiv a_m/v_m$, where $v_m \simeq 50\; \mu {\rm {m/s}}$ is the mean swimming speed of isolated mbo2 cells. We  observe that there are no well-developed plateaux in the plots of the MSD, $F_s(k,t)$, and  $Q(t)$; such plateaux 
are desiderata for an active glass ~\cite{sadhukhan2024perspective}. However, the presence of a peak in the four-point correlation function $\chi_4(t)$ suggests dynamic heterogeneity in our algal systems.
In Appendix~\ref{appendix:active_glass}, we show the variation, with respect to the average cell-density, of the intermediate scattering function, the overlap function, the 4-point correlation function and the MSD; and
how the intermediate scattering function varies for different values of $k$. We note that the MSD plots show diffusive behaviors at short time scales, which is typical of crowding in dense suspensions~\cite{Koorehdavoudi2017-lz}; at longer time scales, the MSD show super-diffusive behavior, as expected~\cite{Ariel2015-dc,mukherjee2021anomalous}.

\section{Discussion}

In summary, we have demonstrated that dense suspensions of Chlamydomonas reinhardtii algal cells display a new type of turbulence-like dynamics with strongly non-Gaussian velocity distributions, a feature distinct from both bacterial and classical fluid turbulence ~\cite{pandit2017overview,boffetta2012two,dunkel2013fluid,alert2020universal,mukherjee2023intermittency,KiranPRF2023}  . This spatiotemporal chaotic state also displays power-law energy spectra and small-scale intermittency that are quantitatively different from their counterparts in conventional fluid turbulence and in bacterial turbulence . Although the overall dynamics in algal suspensions slows down with increasing number density of cells, we do not observe a cross over to a state that can be identified as a \textit{bona fide}  active algal glass. It might well turn out that an active algal glass appears at densities values larger than those we have considered here; however, such high densities are hard to achieve in experiments, so the explorations of active glassy dynamics in very dense suspensions of \textit{C. reinhardtii} remains a challenge for future studies.

Furthermore, our observations have obvious biological implications; such spatiotemporal chaos enhances mixing and transport, so we conjecture that it aids \textit{C. reinhardtii} in foraging for and outrunning  diffusing nutrient molecules~\cite{debasmita_flowinversion}. Microbial turbulence is often associated with the presence of topological defects and orientational instabilities{~\cite{pandit2025particles,giomi2015geometry,Yashunsky2024-hr}. Our study shows that turbulence-like dynamics can also arise in the absence of orientational instabilities. Hence, we hypothesize that active chaotic flow is a universal feature of free-swimming micro-organisms, regardless of their alignment interactions. 

It is worth noting that many systems exhibit low-Reynolds number turbulence. For example, elastic turbulence in fluids with polymer additives~\cite{groisman2000elastic,gupta2017melting,Steinberg2021elastic}, binary-fluid mixtures with low interfacial tensions~\cite{padhan2024interface}, or active fluids like bacterial suspensions~\cite{alert2021,pandit2025particles,wensink2012meso,dunkel2013fluid}. In all these cases, there is another control parameter that induces turbulence, even though the Reynolds number is low. In the case of elastic turbulence, this is the Weissenberg number, for interface-induced turbulence this is the Capillary number (inversely proportional to the surface tension), and in active fluids it is a dimensionless measure of the activity. The word turbulence, commonly adopted in the active-matter and bacterial- suspension literature [see, e.g., Refs.~\cite{bratanov2015new, wensink2012meso, alert2021, padhan2024novel, pandit2025particles}], is used to denotes a spatiotemporally chaotic, multi-scale flow state characterized by broadband spectra, vortex structures, and strong non-Gaussian velocity statistics. Inertial cascades might exist here, but to analyze them systematically requires a hydrodynamical model. 

Current theoretical models of active turbulence are either inappropriate or insufficient to explain our results. The Toner-Tu-Swift-Hohenberg (TTSH) model~\cite{bratanov2015new, wensink2012meso,alert2021, pandit2025particles}, which is commonly used for bacterial turbulence, cannot be applied to algal system because the \textit{C. reinhardtii} cells do not swarm or flock; and they have an aspect ratio very close to one (unlike rod-like bacteria). Recent work on theoretical and numerical study of the active Cahn Hilliard Navier Stokes (CHNS) equations suggests that active turbulence might be realized in dense suspensions of active-scalar matter~\cite{padhan2024novel}. However, this active-CHNS model also does not seem to be adequate for the description of our algal system, insofar as the statistical properties are different from those in the model. Therefore, the theoretical modeling of spatiotemporal chaos in dense suspensions of C. reinhardtii cells remains a challenging open problem, whose resolution could lead to scaling arguments for the spectral exponents and a theoretical understanding of the statistics of velocity increments.

\section{Materials and methods}

\subsection{Cell culture}
\label{subsec:cell-culture}

\textit{Chlamydomonas reinhardtii} cells of both the WT and mbo2 strains are inoculated from agar plates into TAP+P media and grown at 25$\degree$ Celsius in 12:12 hour day-night cycles inside an orbital shaker at 137rpm~\cite{sujeet_phototaxis}. When the cells are in the logarithmic phase of their growth cycle, the culture is collected in Eppendorf tubes and then centrifuged multiple times (at 0.3rcf for 4 minutes) to obtain samples with different cell concentrations.\\

\textit{C. reinhardtii} is a photosynthetic alga that feeds on dissolved inorganic ions or molecules that are in the TAP+P media; these include phosphate and ammonium ions and carbon dioxide from the surrounding fluid; light is the main source of energy~\cite{Tam2011-rz,Kiørboe2008amechanical}. The macronutrients that limit algal metabolism and growth are nitrogen and carbon~\cite{Khan2018-vm,Short2006-dj,Kiørboe2008amechanical}. 
Flow fields  that are flagella-driven help to distribute these dissolved solute molecules uniformly through fluid mixing and transport~\cite{Kiørboe2008amechanical,Tam2011-rz,Ding2014-xx,Short2006-dj,leptos2009dynamics,gollub2011enhancement}.  


\subsection{Surface modification of glass slides, cover slips, and beads}
\label{subsec:coating}

A polyacrylamide brush is coated on the glass slides and cover slips to avoid non-specific adhesion of cells and beads on the glass surfaces~\cite{debasmita_flagella}. Also, the microspheres (Sulphate latex 200nm beads, Thermofisher) are coated with PLL-PEG to impart steric stabilization, thereby reducing inter-particle aggregation and obtaining mono-dispersed microspheres in the media~\cite{debasmita_flagella}.

Despite the surface functionalization, a small fraction of cells get stuck on the glass surface. Samples of the same cell-type and concentration typically have a percentage of stuck cells that varies between $1\%$ and $4\%$. This small proportion of stuck cells does not play a significant role in the measured statistics of the system.

\subsection{Microscopic imaging}
\label{subsec:imaging}

Double-sided tape of height 10 micron (Nitto Denko Corporation) is used as a spacer between the coverslip and the glass slide to obtain the chambers in which the cell-culture is placed
(Appendix~\ref{height_section}). The sample is kept under an inverted microscope (Olympus IX83) coupled with a CMOS high-speed camera (Phantom Miro C110, Vision Research, Pixel size = 5.6 micron). Different concentrations of both the WT and mbo2 mutant cells are imaged under red-light ($>610$nm) using a 20x objective in bright field at 100 fps.  

The beat-averaged flow fields are shown for isolated algal cells. To measure such flow fields, we take very dilute cell suspensions (with only one cell in the field of view), for which there is no hydrodynamic interference from neighboring cells. These dilute cultures are mixed with 200nm microspheres and then imaged using a 60x phase objective at 500 fps to obtain flow fields of isolated cells ~\cite{debasmita_flowinversion}.

\subsection{Obtaining Velocity vectors, Cell Density, and Flow-fields from the images}
\label{subsec:image_analysis}

The velocity vectors are obtained by tracking the individual cells between successive frames using standard MATLAB tracking routines ~\cite{blair2008matlab}. For dense suspensions, frames are skipped so that there is sufficient displacement between successive frames and the tracking is accurate. To compute the energy spectra and the Okubo-Weiss parameter, the velocity vectors (obtained from discrete individual cells) are interpolated over a finer equally-spaced grid via Natural-neighbor interpolation. For all other quantities that we compute, we use the non-interpolated velocity vectors.

The spatial cell density is determined by calculating the area fraction covered by the cells at a particular location for each frame 
(Appendix~\ref{density_section}). 
To compute the beat-averaged flow fields of isolated WT and mbo2 cells, the individual microspheres are tracked and their velocity vectors are computed~\cite{debasmita_flowinversion}. For the mbo2 cells, the flow-field data were averaged over 2820 frames or approximately 282 flagellar beat cycles. For the WT cells, the flow-field data were averaged over 170 frames or approximately 17 flagellar beat cycles.

\subsection{Statistical measures of dense algal suspensions}
\label{subsec:spectra}

The statistical properties that we use to explore spatiotemporal chaos in our system are given below:

\begin{itemize}
\item The instantaneous concentration and energy spectra are, respectively:
\begin{eqnarray}
\Phi(k,t) &=& \frac{1}{2} \sum_{k-\frac{1}{2}\le k^{'}\le k+\frac{1}{2}} [\hat{\rho}(\boldsymbol{k}',t)\hat{\rho}(-\boldsymbol{k}',t)]\,;\label{eq:Skt}\\
E(k,t) &=& \frac{1}{2} \sum_{k-\frac{1}{2}\le k^{'}\le k+\frac{1}{2}} [\hat{\boldsymbol{u}}(\boldsymbol{k}',t)]\cdot[\hat{\boldsymbol{u}}(-\boldsymbol{k}',t)]\,;\label{eq:Ekt} 
%
\end{eqnarray}
here, carets denote spatial Fourier transforms, and $k$
and $k'$ are the moduli of the wave vectors $\boldsymbol{k}$ and $\boldsymbol{k}'$. 
\item In the statistically
steady state, we obtain the averaged spectra
\begin{equation}
    \Phi(k) = \langle \Phi(k,t) \rangle_t\,;\;\; E(k) = \langle E(k,t) \rangle_t\,;
    \label{eq:Avspect}
\end{equation}
here, $\langle \cdot \rangle_t$ is the time average; typically we take the mean over $\simeq 100$ energy spectra and $\simeq 400$ concentration spectra, obtained from configurations of the velocity and density fields, at well-separated times in the nonequilibrium statistically steady state. We also average over different samples with similar densities [Fig.~\ref{fig:spectra}].

\item The scale-dependent energy flux $\Pi(k)$ is defined as follows:
\begin{eqnarray}
    T^u(k)&=&\sum_{k'= k-1/2}^{k'= k+ 1/2} \langle\widetilde{\bm{u}(-\bm{k}'}).\bm{P}(\bm{k}').\widetilde{(\bm{u}.    \nabla\bm{u})}(\bm{k}')\rangle_t \,;\nonumber  \\
    \Pi(k)&=&\sum_{k'=0}^{k'=k} T^u(k')\,; 
    \label{eq:Tketc}
\end{eqnarray}
here, the transverse projector $\bm{P}(\bm{k})$ has the components $P_{ij}(\bm{k})=[\delta_{ij} -(k_ik_j/k^2)]$.

\item For various PDFs, we also compute the $x$-component of the velocity field $\boldsymbol{u}$ and the longitudinal velocity increments~\cite{pandit2009statistical} 
\begin{equation}
\delta \boldsymbol{u}_{\parallel}({\boldsymbol{x}},{l},t) \equiv [{\boldsymbol{u}}({\boldsymbol{x}}+{\boldsymbol{l}},t) - {\boldsymbol{u}}({\boldsymbol{x}},t)] \cdot \frac{{\boldsymbol{l}}}{l}\,,
\label{eq:velinc}
\end{equation} 
\item Furthermore, we compute the Okubo-Weiss parameter~\cite{weiss1991dynamics,okubo1970horizontal,pandit2017overview}
\begin{eqnarray}
    \Lambda &=&\Omega^2\ - D^2;\;\;
    D = \frac{\nabla \bm u + \nabla \bm u^T}{2}\,;\nonumber\\
    \Omega &=& \frac{\nabla \bm u - \nabla \bm u^T}{2}\,;
    \label{eq:Okubo}
\end{eqnarray}
$D$ and $\Omega$ are the rate-of-deformation and rate-of-rotation tensors, respectively;
$\Lambda > 0$ in vortical regions of the flow and $\Lambda < 0$ in  strain-dominated regions. This Okubo-Weiss criterion was originally developed for inviscid two-dimensional flows. However, this criterion provides a useful measure of flow properties even for viscous 2D 
flows ~\cite{rivera2001universal, Perlekar2009-kk,Lemma2019-gv}.
\item We compute the flatness $F_4(l)$ to check for intermittency in our system ~\cite{Frisch-CUP,perlekar2006manifestations,sahoo2011systematics,pandit2017overview}:
\begin{eqnarray}
    F_4(l) &=& \frac{S_4(l)}{(S_2(l))^2} ;\nonumber\\ 
    S_p(l) &=& \langle [\delta \boldsymbol{u}_{\parallel}({\boldsymbol{x}},{l},t)]^p \rangle_{{\boldsymbol{x}},t}\,;
    \label{eq:flat}
\end{eqnarray}
here, $S_p(l)$ is the longitudinal-velocity structure function and $\langle \cdot \rangle_{{\boldsymbol{x}},t}$ denotes the average over the spatial origin ${\boldsymbol{x}}$ and the time $t$.

\end{itemize}
\vspace{0.5cm}

\subsection{Statistical measures for active glasses}

In the active-glass literature~\cite{sadhukhan2024perspective}, various statistical properties are used to characterise a glass. 
The analysis that we present here is based on the particles (here, \textit{C. reinhardtii} cells)  that comprise the glass.
The statistical properties that we use to explore the formation of an active glass
in our system are given below:
\begin{itemize}
\item The mean-square displacement (MSD)
\begin{eqnarray}
    <\Delta x^{2}(t)> &=& \langle \frac{1}{N} \sum_{i=1}^{N} [\bm{x}_i(t+t_0)-\boldsymbol{x}_i(t_0)]^{2} \rangle_{t_0}\,;
    \label{eq:msd}
\end{eqnarray}
here, $\bm{x}_i(t)$ is the position of the $i^{th}$ cell at time $t$ and $N$ is the total number of cells. We use $N \simeq 1000$.
\item The self intermediate scattering function $F_s(\boldsymbol{k},t)$, at wave vector $\boldsymbol{k}$ and time $t$,
\begin{eqnarray}
    F_s(\boldsymbol{k},t) &=& \langle \frac{1}{N} \sum_{i=1}^{N} e^{i\boldsymbol{k}.(\boldsymbol{x}_i(t+t_0)-\boldsymbol{x}_i(t_0))} \rangle_{t_0}\,;
    \label{eq:fskt}
\end{eqnarray}
in our calculations we use $\boldsymbol{k} = k_x \hat{\bm{x}}$.
\item The overlap function 
\begin{eqnarray}
    \tilde{Q}_i &=& \langle W(a-|\boldsymbol{x}_i(t+t_0)-\boldsymbol{x}_i(t_0)|) \rangle_{t_0}\,;\nonumber\\
    Q(t) &=& \frac{1}{N} \sum_{i=1}^{N} \tilde{Q}_i \equiv \langle \tilde{Q}_i \rangle_{i};
    \label{eq:qt}
\end{eqnarray} 
where $W$ is the Heaviside step function and the length scale $a$ is taken as the typical vibrational amplitude of the glass particles; in our systems, we chose $a$ such that $a/a_w = 0.18$ for the WT cells and $a/a_m = 0.097$ for the mbo2 mutant.
\item The four-point correlation function
\begin{eqnarray}
    \chi_4(t) &=& N[\langle \tilde{Q}_i^{2} \rangle_{i} - [\langle \tilde{Q}_i \rangle_{i}]^{2}]\,;
    \label{eq:chi4t}
\end{eqnarray}
\end{itemize}

\section{Supplementary material}

Video captions:
\begin{itemize}

\item
V1: This experimental video shows the spatiotemporal evolution of WT cell suspensions with $\bar \rho = 0.50$. The video is captured with a Phantom Miro C110 Camera using a 20x bright field objective in an Olympus IX83 microscope. \\
\item
V2: This experimental video shows the spatiotemporal evolution of WT cell suspensions with $\bar \rho = 0.74$. The video is captured with a Phantom Miro C110 Camera using a 20x bright field objective in an Olympus IX83 microscope.\\
\item
V3: This experimental video shows the spatiotemporal evolution of mbo2 cell suspensions with $\bar \rho = 0.50$. The video is captured with a Phantom Miro C110 Camera using a 20x bright field objective in an Olympus IX83 microscope. \\
\item
V4: This experimental video shows the spatiotemporal evolution of mbo2 cell suspensions with $\bar \rho = 0.75$. The video is captured with a Phantom Miro C110 Camera using a 20x bright field objective in an Olympus IX83 microscope. \\
\item
V5: This video shows the spatiotemporal evolution of the pseudocolor plots of the density field for the WT cells ($\bar \rho = 0.50$).\\
\item
V6: This video shows the spatiotemporal evolution of the pseudocolor plots of the the vorticity $\omega$ field for the WT cells ($\bar \rho = 0.50$). The arrows indicate the velocity vectors; these vectors are scaled by a factor of 10 from their actual displacements for better clarity.\\
\item
V7: This video shows the spatiotemporal evolution of the velocity vectors overlaid on the images of WT cell suspensions with $\bar \rho = 0.50$.\\

\end{itemize}

}

\begin{acknowledgments}
P.V.B and P.S thank Debasmita Mondal for sharing the PTV codes; Tejas G Murthy, Aparna Baskaran, Sriram Ramaswamy and Chandan Dasgupta for their valuable discussions; the ANRF grant CRG/2022/000724 and the CSIR grant 03WS(008)/2023-24/EMR-II/ASPIRE for providing the support to carry out the experiments.
N.B.P, B.M, and R.P thank K.V. Kiran for discussions, the Anusandhan National Research Foundation (ANRF), the Science and Engineering Research Board (SERB), and the National Supercomputing Mission (NSM), India, for support,  and the Supercomputer Education and Research Centre (IISc), for computational resources.

\end{acknowledgments}

\appendix

\begin{figure}[h!]
    \centering
    \includegraphics[width=0.9\linewidth]{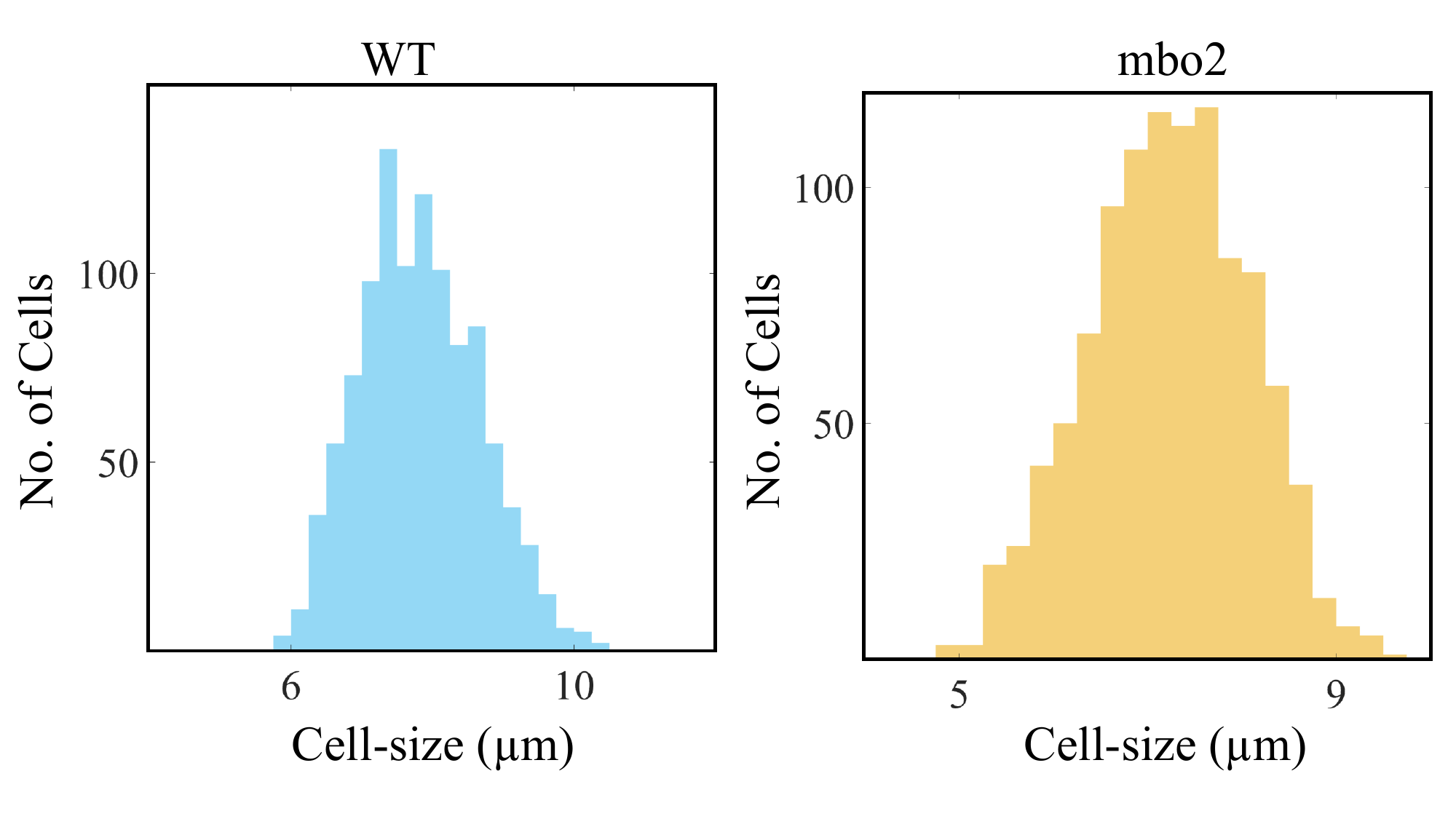}
    \caption{Histogram plots of the cell-sizes for the wild-type cells (left) and the mbo2 mutant(right)}
    \label{cellsize}
\end{figure}

\section{Cell sizes of different strains}
\label{cell_size_section}
Figure~\ref{cellsize} presents histograms of the mean diameters for both the Wild-type (WT) and the mbo2 mutant \textit{C. reinhardtii} cells. The average diameter of the WT cell is $a_w = 7.82\pm 0.84\ \mu$m and for
the mbo2 cell $a_m =  7.22\pm 0.85\,\mu$m.

\section{Height measurement of the chamber}
\label{height_section}
We take a dilute solution of microspheres (Sulphate latex 200nm beads, from Thermofisher) on the coverslip and the glass slide; we then heat the glass slide  and the coverslip gently so that the beads get stuck on both the surfaces. After that, immersion oil is put into the sample chamber. We use a 60x oil-immersion phase objective to focus on the beads and determine the chamber height to be \(10.27\pm 0.61\ \mu \)m.

\section{Orientational order}
\label{appendix:orientational_order}

\begin{figure}[h!]

    \centering
    \includegraphics[width=1\linewidth]{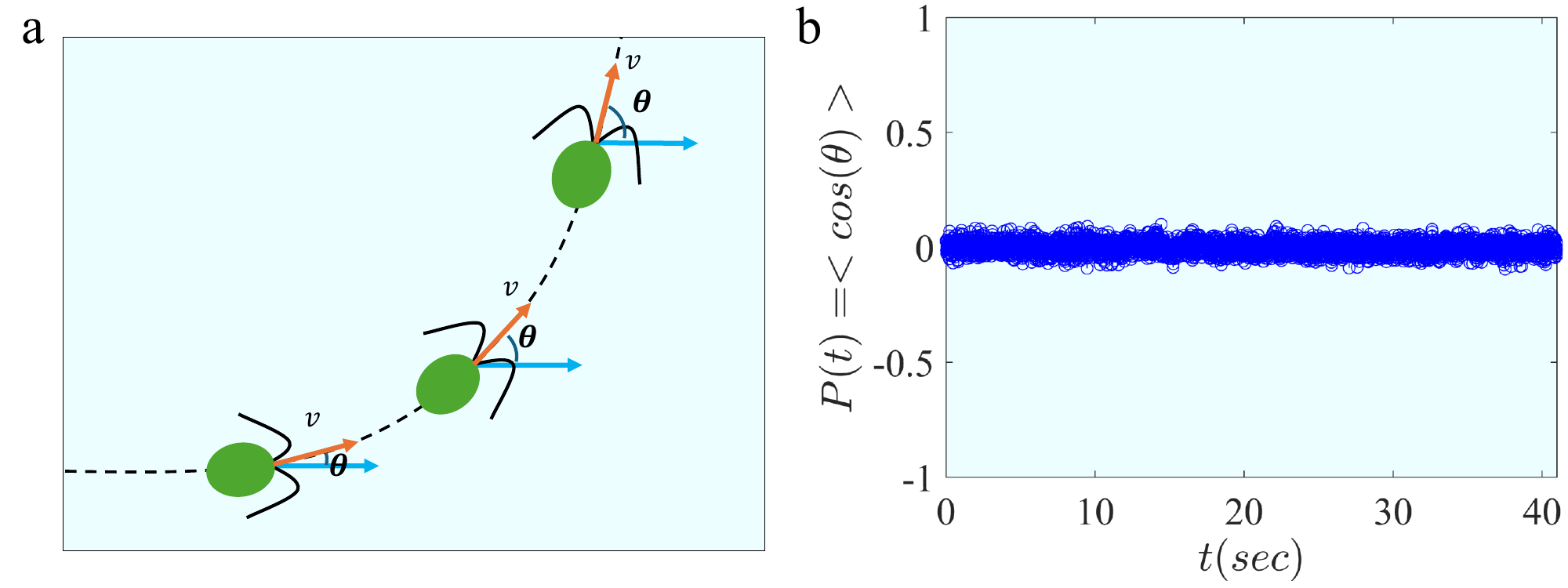}
    \caption{(a) Orientation angle, $\theta$ of a CR cell. (b) Average polarization of the cell suspension as a function of time t, for wild-type (WT) cells and representative $\bar\rho = 0.50$.}
    \label{avg_polar}
\end{figure}

The orientation angle, $\theta$ of a CR cell is the angle of the instantaneous velocity vector of a cell relative to the $x$ axis in the lab frame. Figure~\ref{avg_polar}(a) shows a schematic of the motion of a cell; the orientation angle, $\theta$ and the velocity vector, $v$ are shown along its trajectory. Figure~\ref{avg_polar}(b) shows the average polarization $P(t) = <\cos\theta>$ versus time t.
\begin{figure}[h!]

    \centering
    \includegraphics[width=1\linewidth]{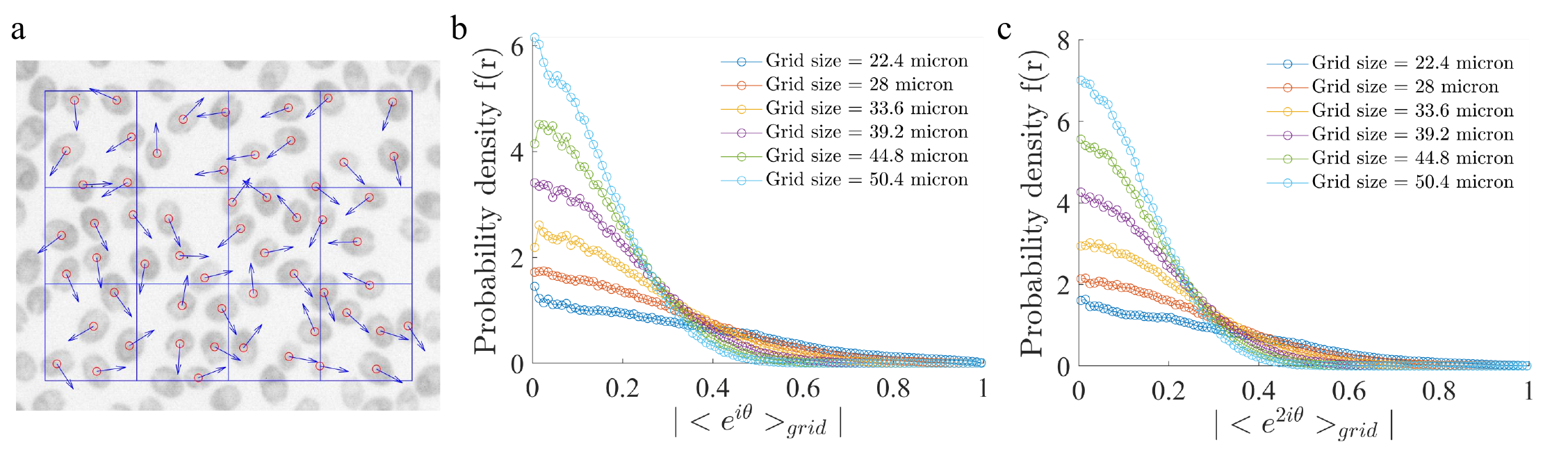}
    \caption{(a) The orientation vectors of the CR cells, over a microscope image with a superimposed spatial grid; (b) and (c) show, respectively, the probability distribution functions (PDFs) $f(r)$ of the orientational order parameter $| < e^{i\theta} >_{grid} |$ and the nematic order parameter $| < e^{2i\theta} >_{grid} |$ averaged over different grid sizes.}
    \label{orientation_pdf}
\end{figure}

\begin{figure}[h!]
    \centering
    \includegraphics[width=1\linewidth]{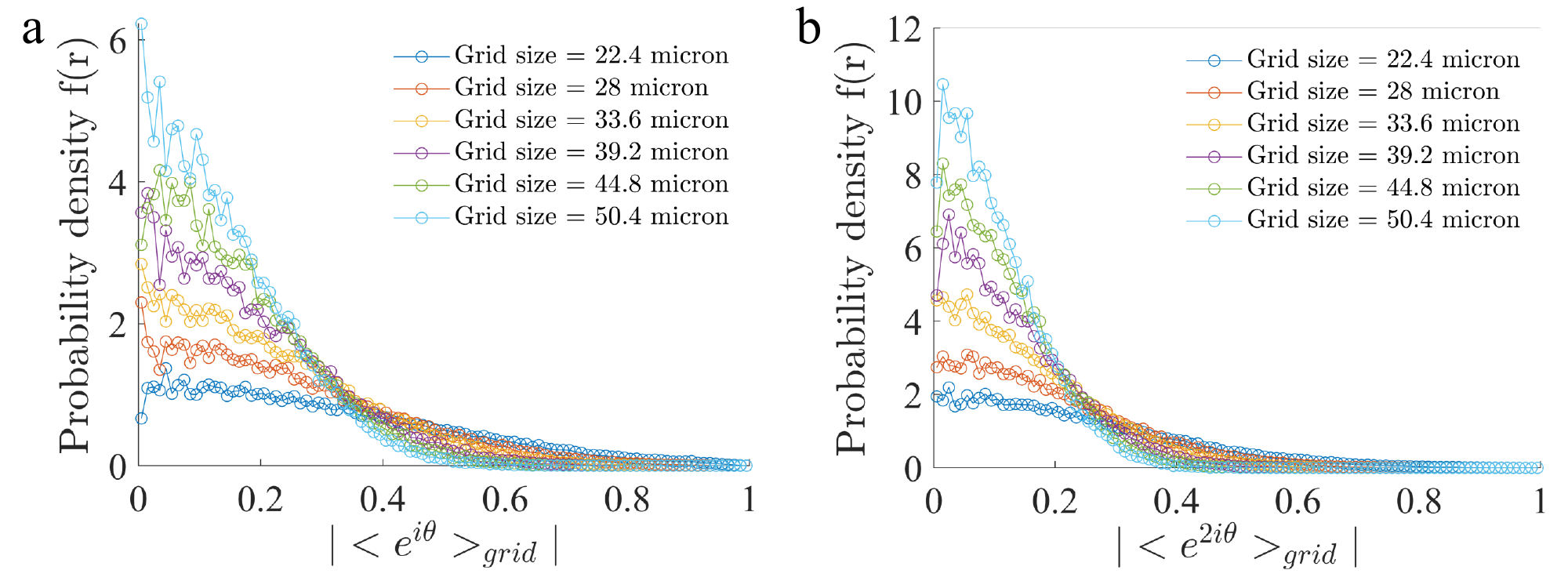}
    \caption{The probability distribution functions (PDFs) $f(r)$ of (a) the orientational order parameter $| < e^{i\theta} >_{grid} |$ and (b) the nematic order parameter $| < e^{2i\theta} >_{grid} |$ for WT cells and $\bar \rho = 0.74$}
    \label{orientation_pdf_high}
\end{figure}
Figure~\ref{orientation_pdf}(a) shows the orientation vectors of the CR cells ($\bar\rho = 0.50$), over a representative image of the cell suspension with a superimposed spatial grid. Figures~\ref{orientation_pdf}(b) and (c) present, respectively, the probability distribution functions (PDFs) of the orientational order parameter $| < e^{i\theta} >_{grid} |$ and its nematic counterpart $|<e^{2i\theta}>_{grid}|$ [Eq.~\eqref{eqn_pdf_fr}]. The subscript $grid$ indicates an average over cells within our grids and we consider several grid sizes. The probability distribution function (PDF) $f(r)$ is defined as 

\begin{equation} \label{eqn_pdf_fr}
    f(| < e^{ki\theta} >_{grid} | = r) = \frac{1}{2\pi r} \int_{0}^{2\pi}  pdf(| < e^{ki\theta} >_{grid} | = r, arg(< e^{ki\theta} >_{grid}) = \phi) d\phi\,.
\end{equation}

Had there been any peaks in the PDFs in Figs.~\ref{orientation_pdf}(b) and (c) they would have signified orientational or nematic order. Similar plots of the probability distribution function (PDF) $f(r)$ for $\bar \rho = 0.74$ is shown in Fig.~\ref{orientation_pdf_high}.

\section{Spatial velocity-velocity correlation}
\label{appendix:velocity_correlation}

\begin{figure}[h!]

    \centering
    \includegraphics[width=1\linewidth]{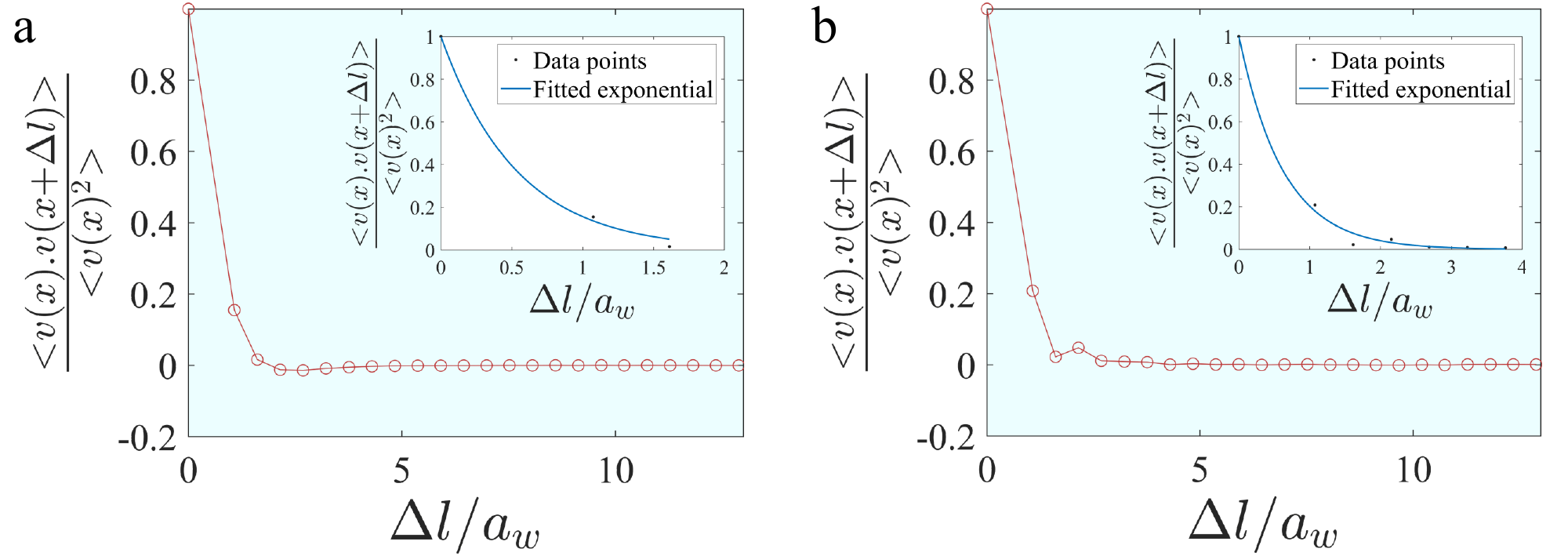}
    \caption{Spatial velocity-velocity correlation function for WT cells versus the scaled distance $\Delta l/a_w$ at two representative values of $\bar\rho$  (a) 0.50 and (b) 0.74. Insets show exponential-decay fits to the initial part of the correlation functions. The decay correlation length $\lambda = 0.54$ for $\bar\rho = 0.50$, and $\lambda = 0.63$ for $\bar\rho = 0.74$, indicating that velocity correlations die out very rapidly (even below one cell-length).}
    \label{spatial_vel_correlation}
\end{figure}

Figure~\ref{spatial_vel_correlation} shows the spatial velocity-velocity correlation function, $\frac{<v(x).v(x+\Delta l)>}{<v(x)^2>}$ for WT cells for two different densities. We fit the initial part of the graph with an exponential function, $\frac{<v(x).v(x+\Delta l)>}{<v(x)^2>} = e^{-(\Delta l/a_w)/\lambda}$ to estimate the correlation length $\lambda$.

\section{Characteristic length scales}
\label{appendix:length_scales}

\begin{figure}[h!]

    \centering
    \includegraphics[width=1\linewidth]{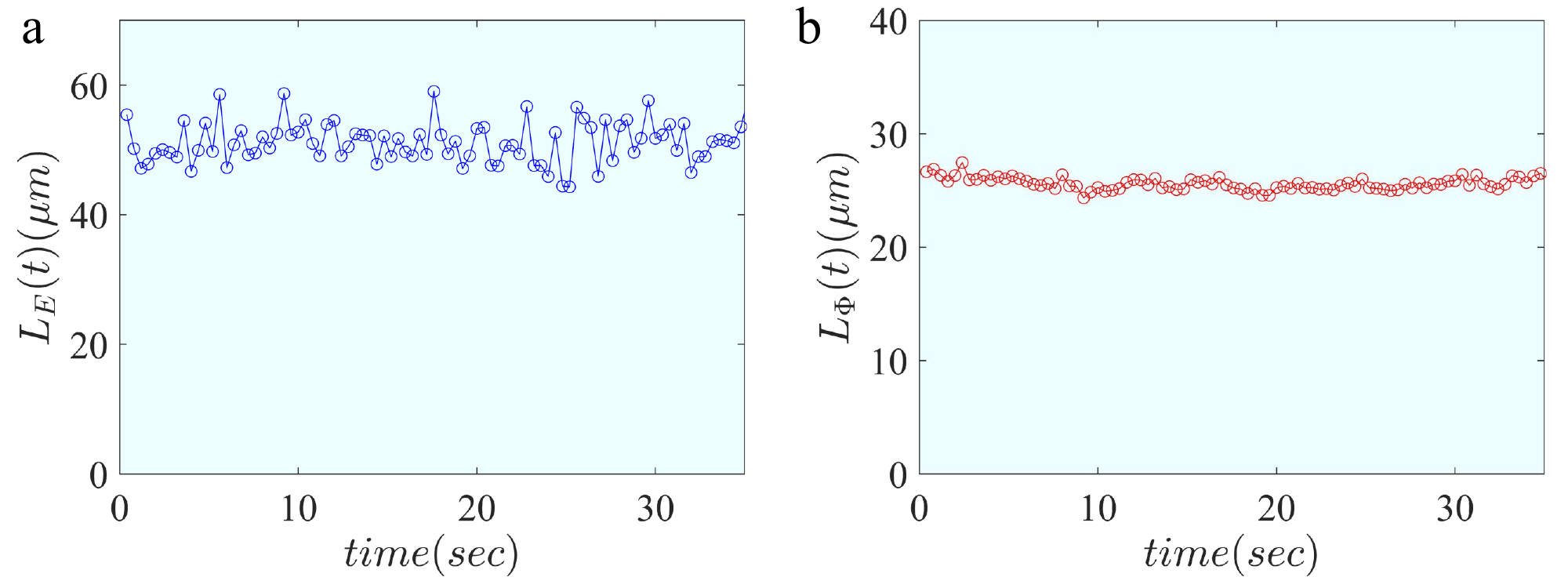}
    \caption{The temporal evolution of the characteristic length scales (a) $L_E$ and  (b) $L_\phi$ for WT cells (for $\bar\rho  = 0.50$). }
    \label{fig:length_scales}
\end{figure}

\begin{figure}[h!]

    \centering
    \includegraphics[width=0.8\linewidth]{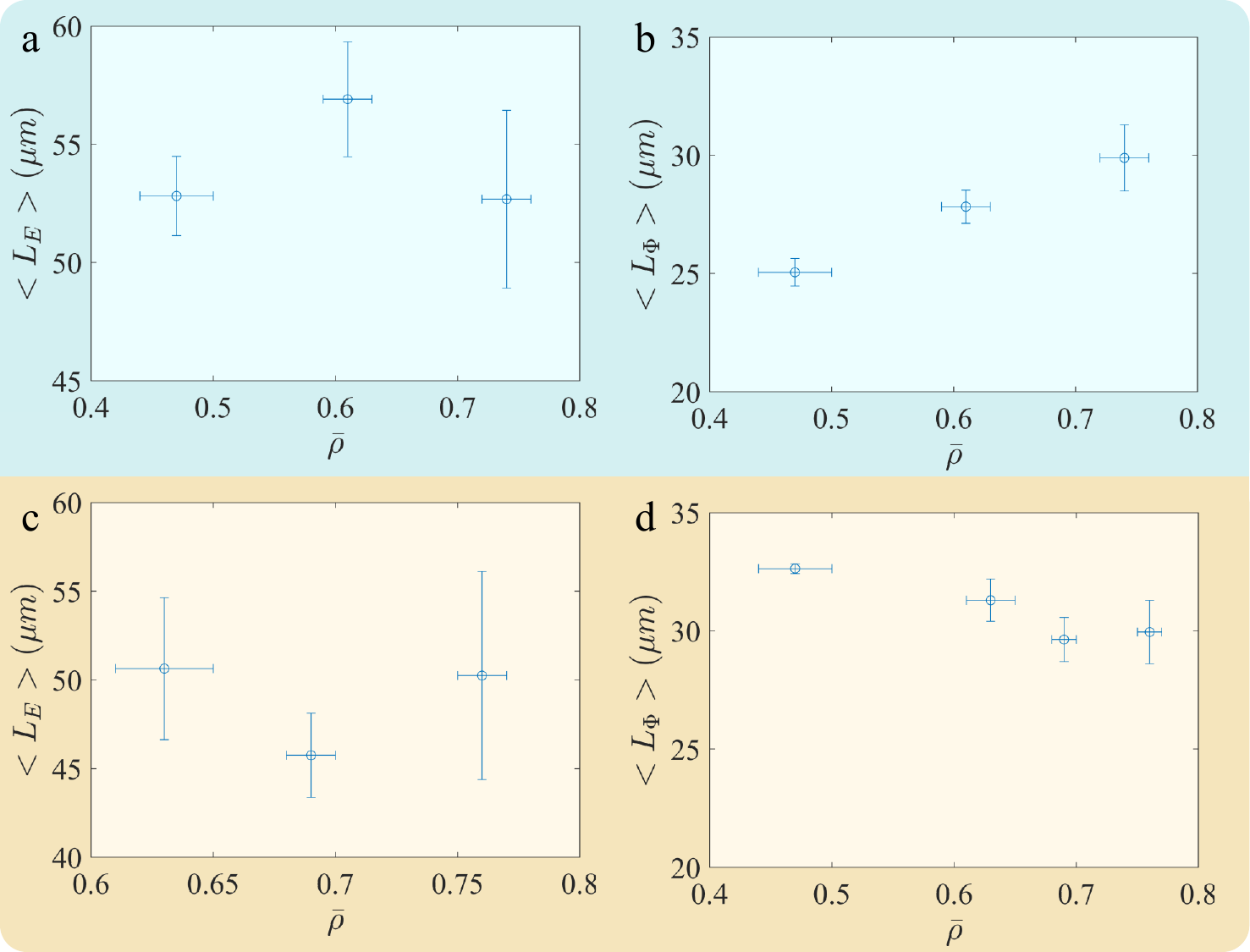}
    \caption{The characteristic length scales (a) $L_E$ and  (b) $L_\phi$ for WT cells of different cell-densities. Panels (c) and (d) are the mbo2 counterparts of (a) and (b), respectively.}
    \label{lengthscales_rho}
\end{figure}

We plot two natural characteristic length scales $L_E$ and $L_\phi$ as a function of time in Fig.~\ref{fig:length_scales}. The length scales for different cell-densities are shown in Fig.~\ref{lengthscales_rho}.

\begin{figure}[h!]
    \centering
    \includegraphics[width=1\linewidth]{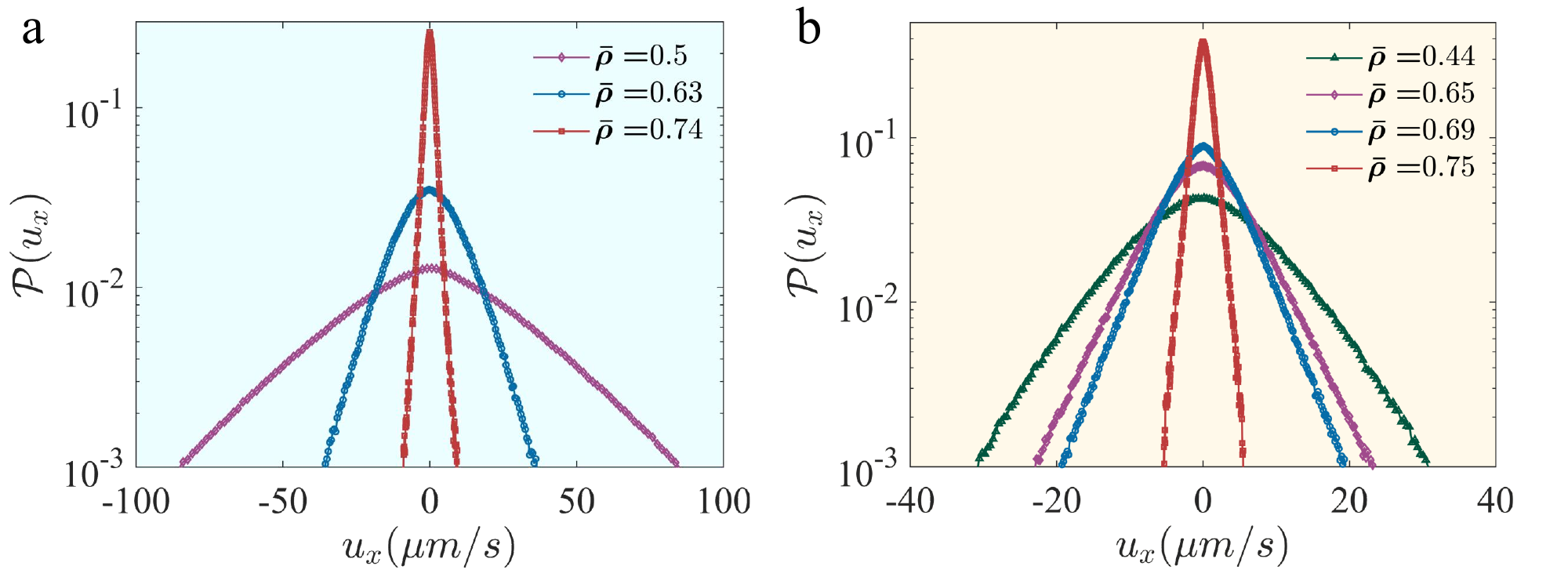}
    \caption{Semi-log plots of the PDFs of the $x$-component of velocity for (a) WT cells (blue background) and (b) mbo2 cells (beige background) showing the flow speeds in $\mu m/sec$. }
    \label{fig:pdf_ux_actual}
\end{figure}

\begin{figure}[h!]
    \centering
    \includegraphics[width=1\linewidth]{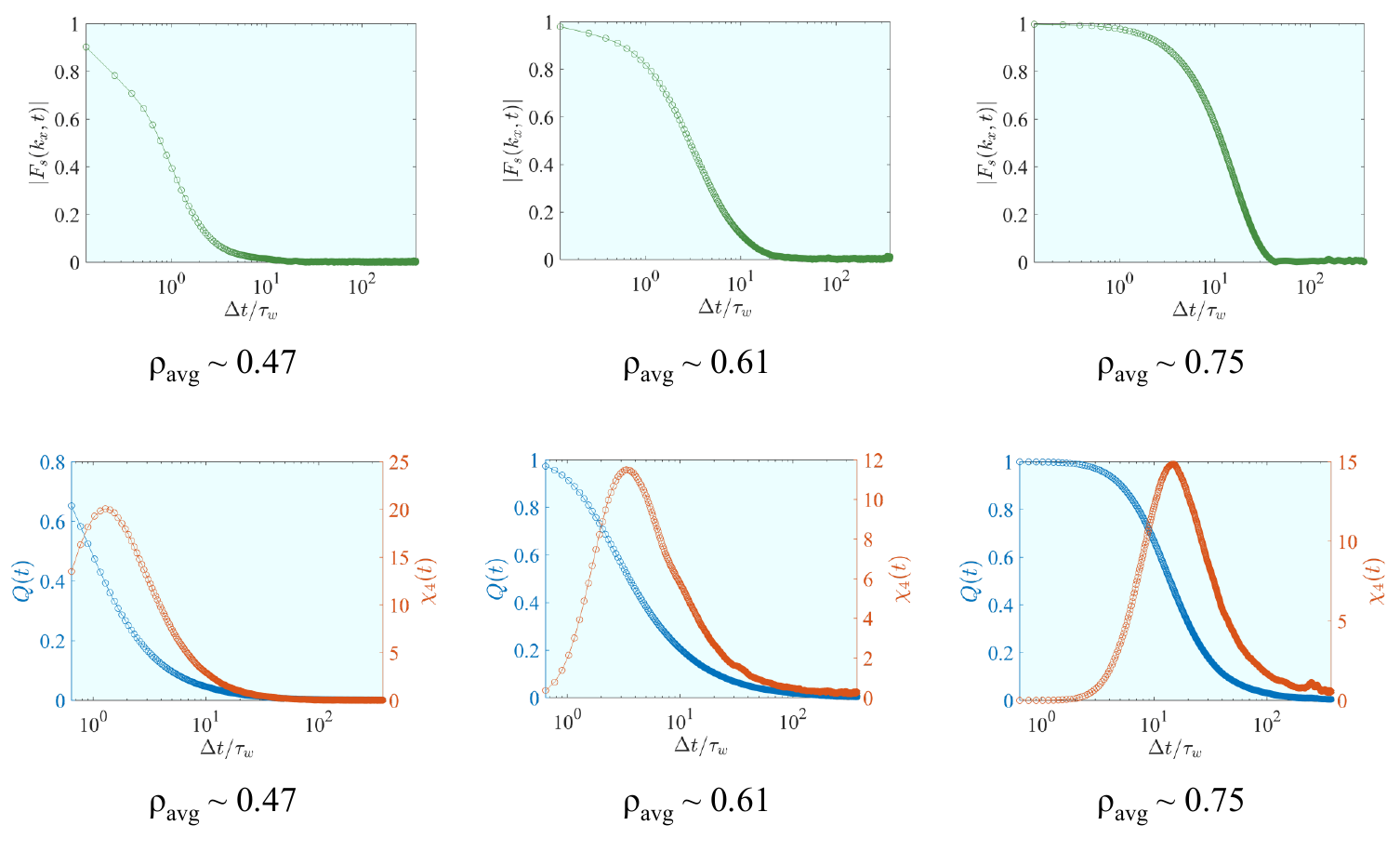}
    \caption{The intermediate scattering function (top panel), the overlap function along with the 4-point correction function (bottom panel) is shown for three different cell densities.}
    \label{fig:fskt_qt_compare}
\end{figure}





\begin{figure}[h!]
    \centering
    \includegraphics[width=0.8\linewidth]{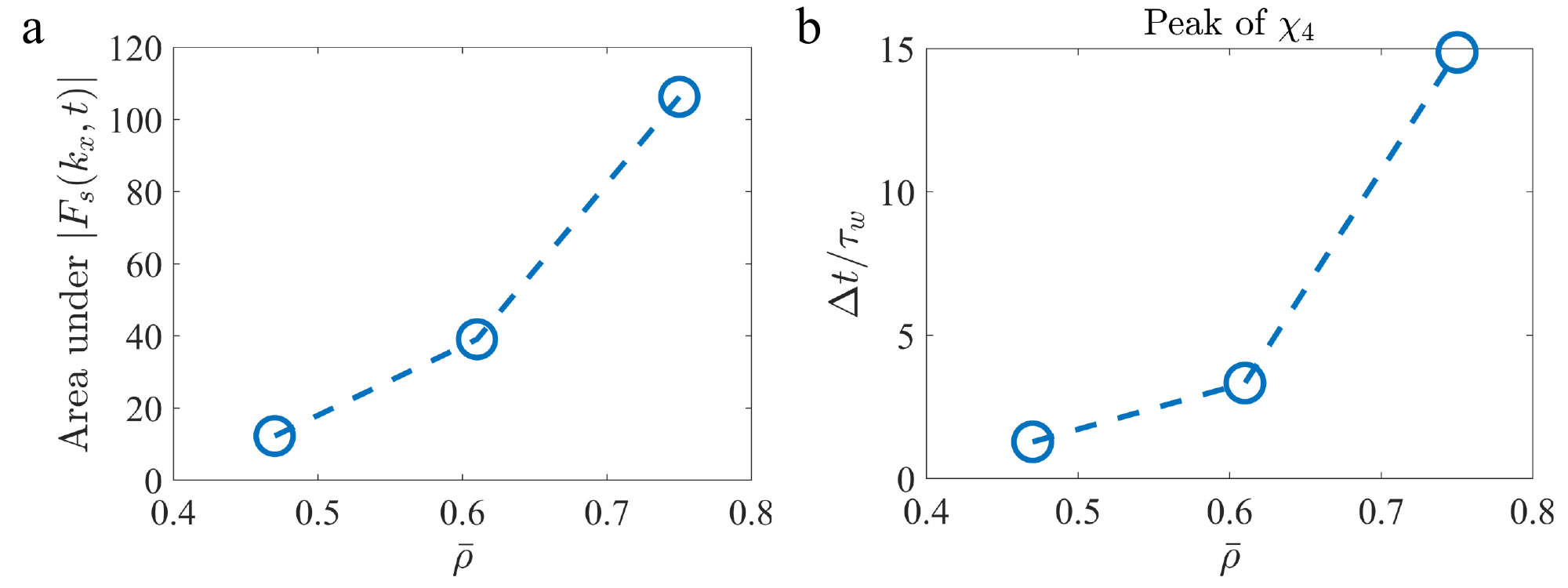}
    \caption{(a) The area under the scattering function and (b) the peak in the 4-point correlation function plotted for three different cell-densities. The dashed line is a guide to the eye.}
    \label{fig:fskt_qt_compare2}
\end{figure}

\begin{figure}[h!]
    \centering
    \includegraphics[width=0.5\linewidth]{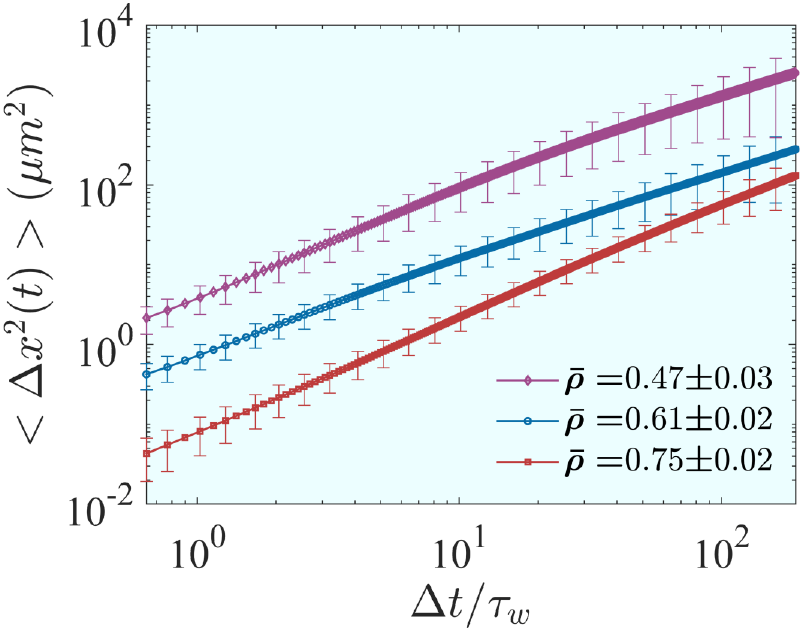}
    \caption{Log-log plots of the mean-square displacement (MSD) for different cell concentrations; the error bars denote $\pm \varsigma(\Delta x^{2}(t))$.}
    \label{fig:msd_rho}
\end{figure}

\begin{figure}[h!]
    \centering
    \includegraphics[width=1\linewidth]{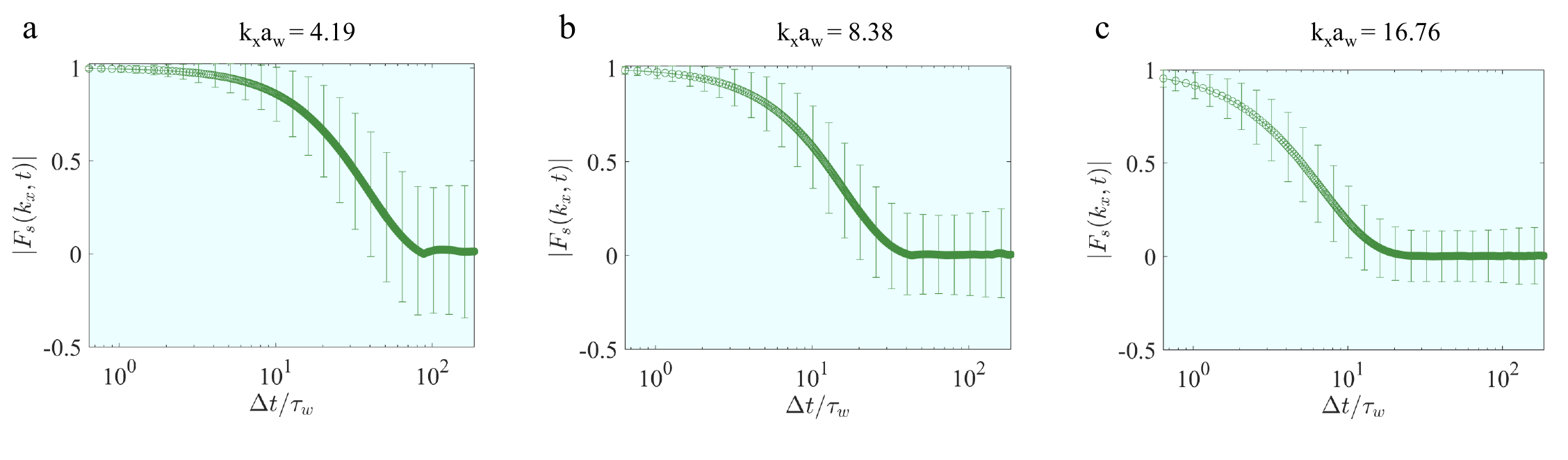}
    \caption{Plots of the intermediate scattering function for different values of $k_x a_w$ (WT cells with $\bar \rho = 0.75 \pm 0.02$).}
    \label{fig:fskt_diff_k}
\end{figure}

\section{PDFs of velocity components}
\label{appendix:PDFvel}

We show the PDFs of the $x$ component of the velocity with the non-normalized flow speeds in Fig.~\ref{fig:pdf_ux_actual}. 

\section{Active-glass characterization}
\label{appendix:active_glass}

The evolution of the intermediate scattering function, the overlap function and the 4-point correlation function with cell-density is shown in Fig.~\ref{fig:fskt_qt_compare}. Figure~\ref{fig:fskt_qt_compare2}(a) shows that the area under the curve of the intermediate scattering function increases with the average cell-density. Similarly, the peak position of the 4-point correlation function in Fig.~\ref{fig:fskt_qt_compare2}(b) increases with an increase in the density. Figure~\ref{fig:msd_rho} shows the mean-square displacement (MSD) for different concentrations of WT cells.

The intermediate scattering function for different values of $k$ are shown in Figure.~\ref{fig:fskt_diff_k}. As $k$ increases, there is a shift in the function. However, the overall behavior of the function remains the same. 

\section{PDFs of velocity increments}
\label{appendix:PDFincrement}

\begin{figure}[h!]
    \centering
    \includegraphics[width=1\linewidth]{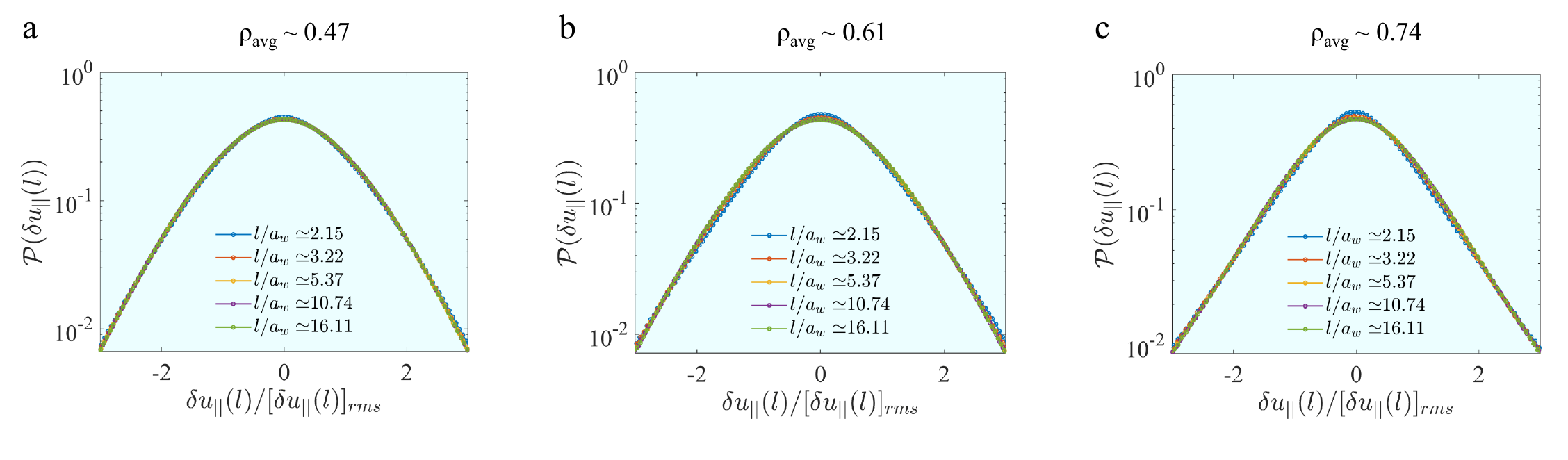}
    \caption{Semilog plots, for WT cells, of the PDFs of longitudinal-velocity increment of the velocity field $\bm{u}$, for different values of separation $l$ and mean density $\bar{\rho}$.}
    \label{fig:vel_inc_lWT}
\end{figure}

\begin{figure}[h!]
    \centering
    \includegraphics[width=1\linewidth]{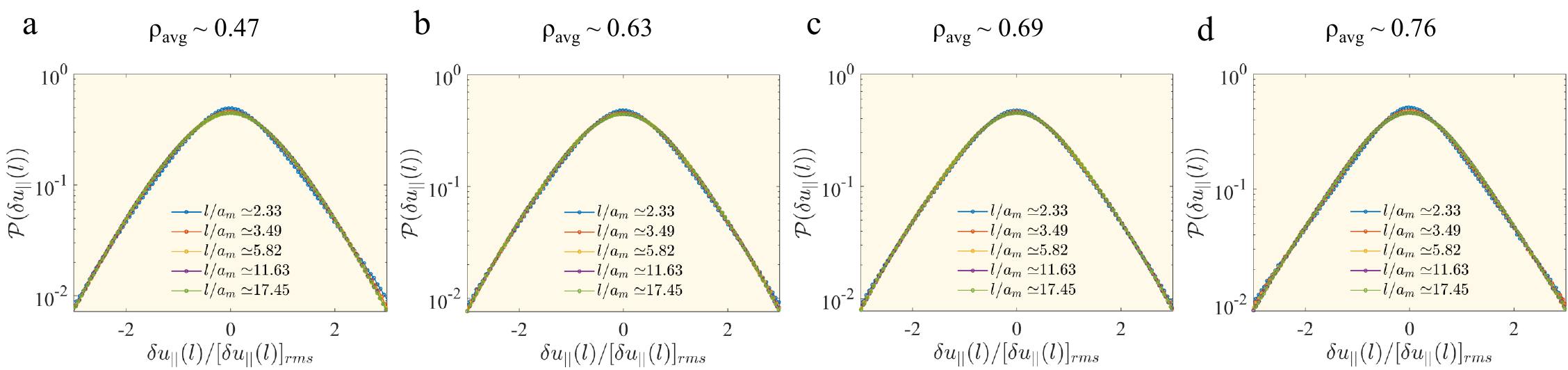}
    \caption{Semilog plots, for mbo2 cells, of the PDFs of longitudinal-velocity increment of the velocity field $\bm{u}$, for different values of separation $l$ and mean density $\bar{\rho}$.}
    \label{fig:vel_inc_lmbo2}
\end{figure}

Figure~\ref{fig:vel_inc_lWT} displays PDFs ($\mathcal{P}$) of the longitudinal velocity increments $\delta \boldsymbol{u}_{\parallel}$(l) of WT cell suspensions for different values of $l$ and $\bar{\rho}$. Figure~\ref{fig:vel_inc_lmbo2} gives the corresponding PDFs for the mbo2 mutant.


\section{Fitted PDFs}
\label{fit_section}
\begin{figure}[h!]

    \centering
    \includegraphics[width=1\linewidth]{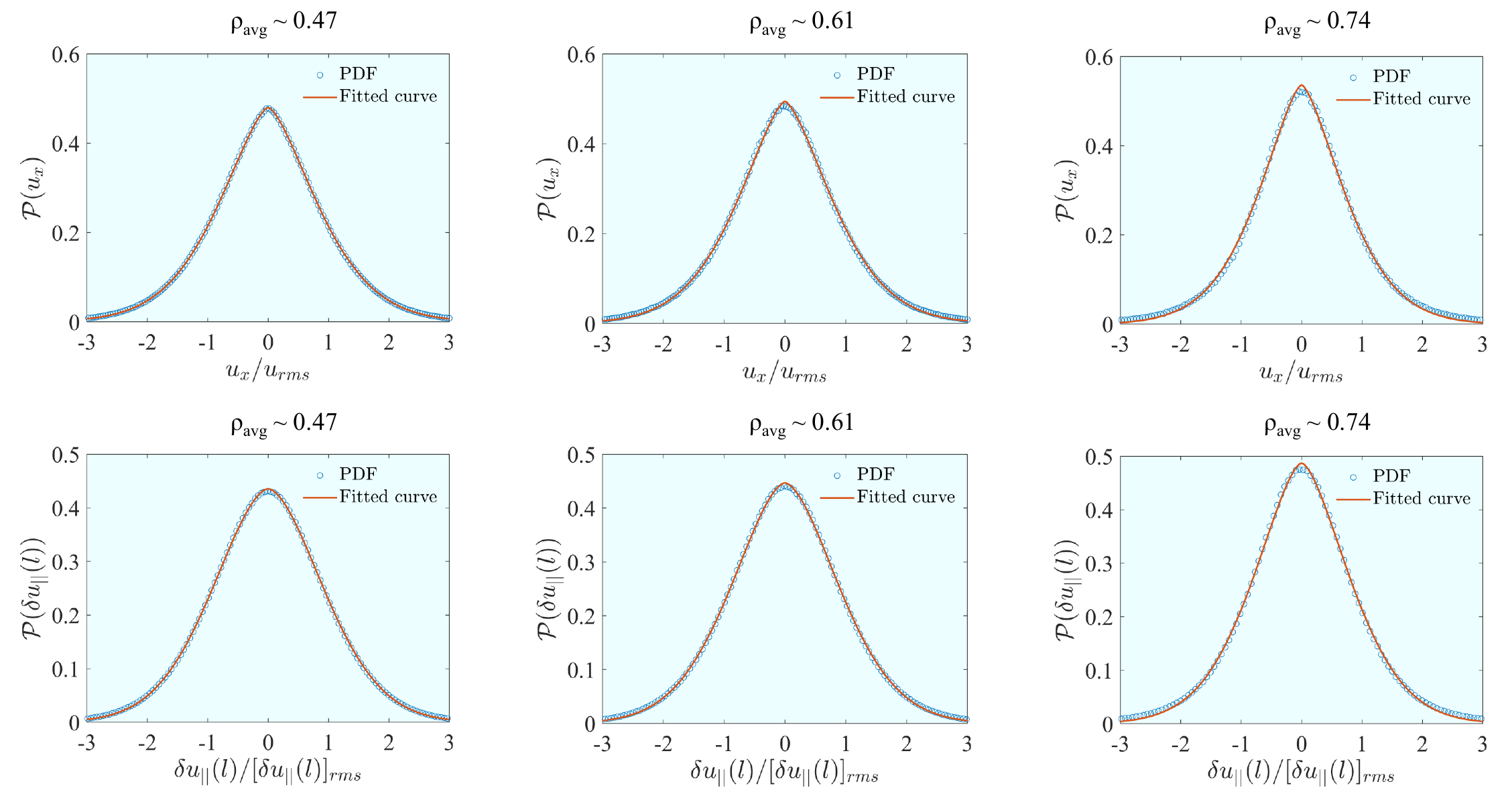}
    \caption{Fitted PDFs of the $x$-component of the velocity (top panel) and the longitudinal velocity increments (bottom panel) for WT cells.}
    \label{fitavg1}
\end{figure}

\begin{figure}[h!]

    \centering
    \includegraphics[width=1\linewidth]{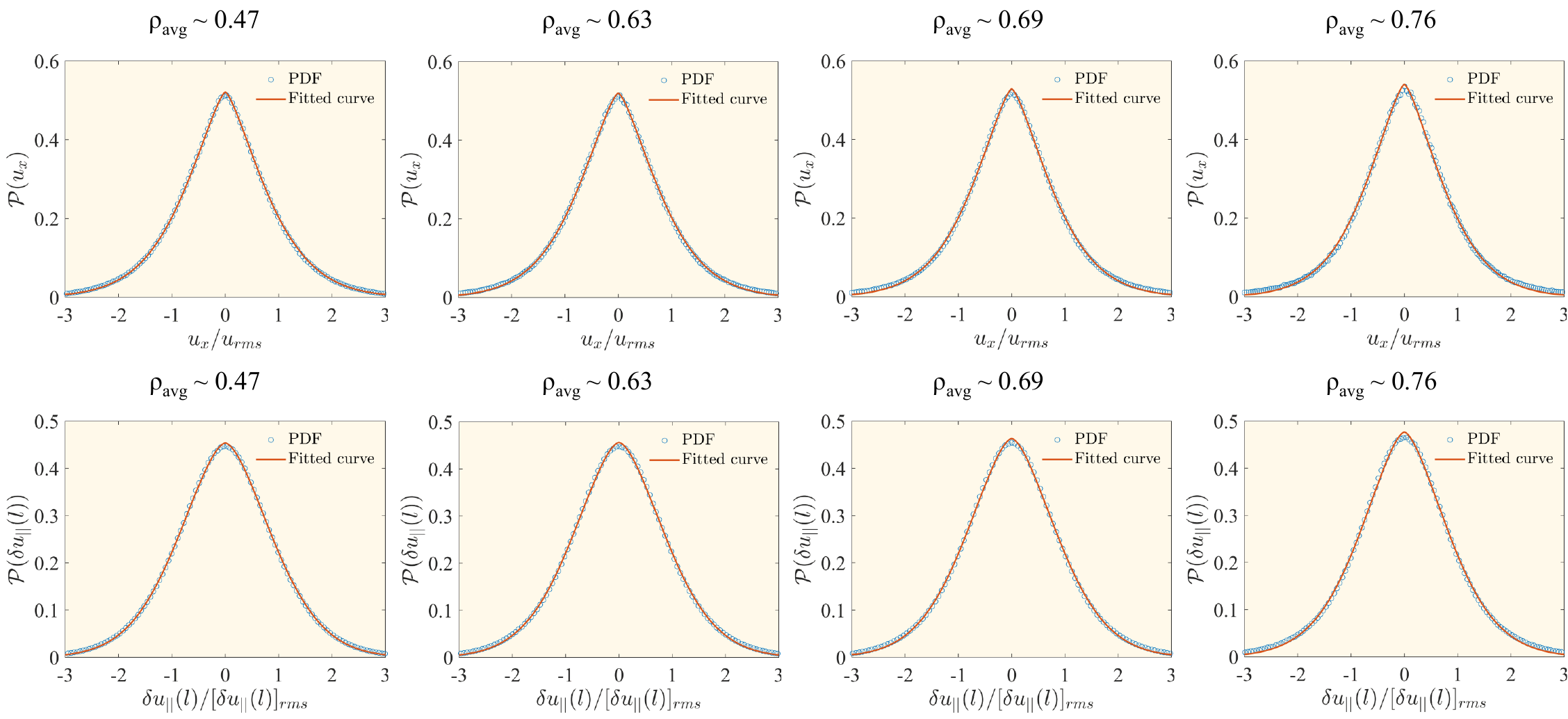}
    \caption{Fitted PDFs of the $x$-component of the velocity (top panel) and the longitudinal velocity increments (bottom panel) for mbo2 cells.}
    \label{fitavg2}
\end{figure}

We find that the PDFs [given in Figs. 4(a), 4(c), 5(a), and 5(c) in the main paper] can be fit to a form of a compressed exponential. Such fits are shown in Fig.~\ref{fitavg1} and Fig.~\ref{fitavg2}. For example, for the WT cells and $\bar \rho = 0.47$, we obtain \\

$\mathcal{P}(\boldsymbol{u}_x) =(0.48)e^{-(0.81)|\boldsymbol{u}_x|^{1.52}}$ and $\mathcal{P}(\delta \boldsymbol{u}_{\parallel}(l)) =(0.44)e^{-(0.64)|\delta \boldsymbol{u}_{\parallel}(l)|^{1.76}}$. 

\begin{figure}
    \centering
   
    \includegraphics[width=1\linewidth]{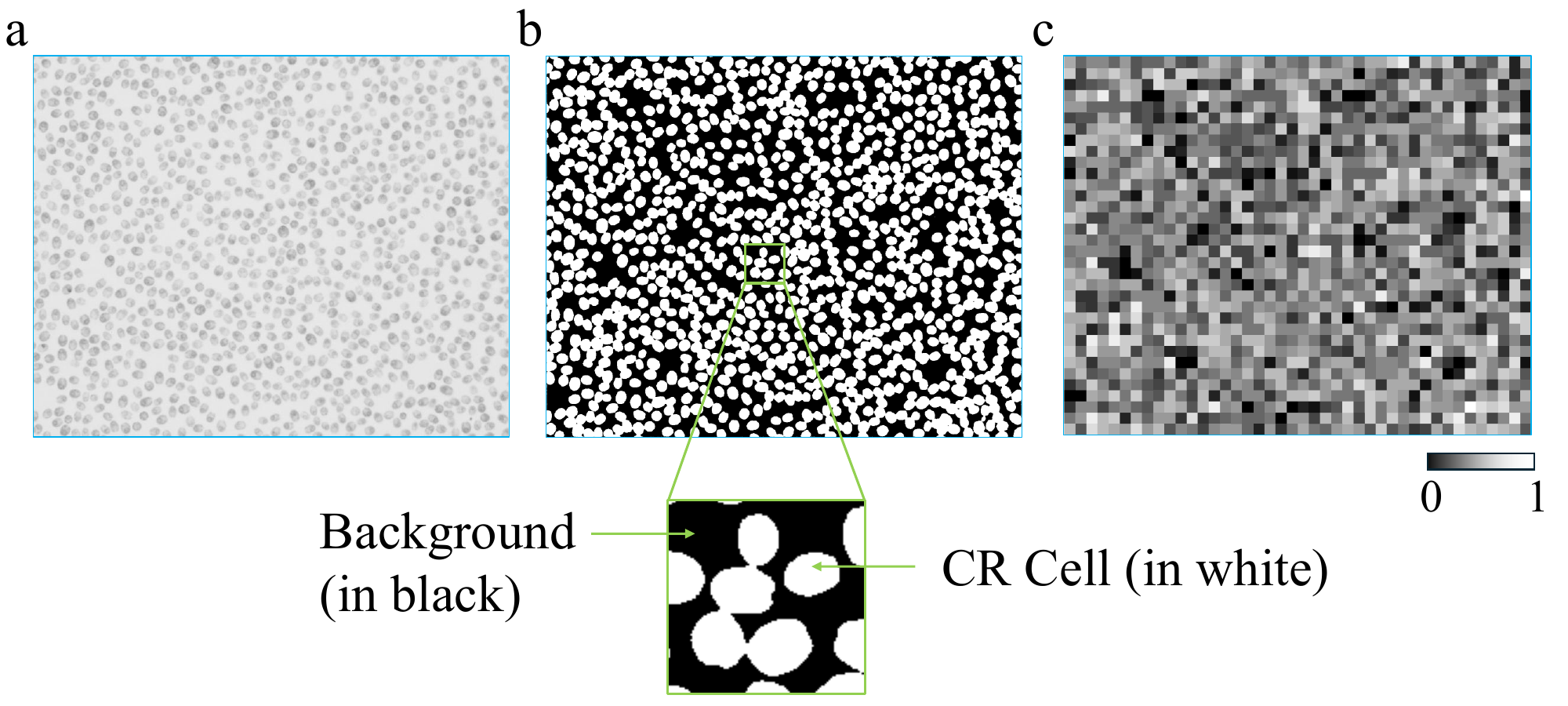}
    \caption{(a) Optical image of a cell suspension. (b) Binarized image where the cells are shown in white and the background in black. (c) A gray-scale plot of the density obtained from the binary image.}
    \label{fig:density}
\end{figure}
\section{Spatial density from the microscopic image}
\label{density_section}

We binarize the image in Fig.~\ref{fig:density} such that the points, where the cells are present, are shown in white (pixel value $= 1$) and the background is black (pixel value $= 0$). To obtain the area fraction covered by the cells in a given region, we count the number of pixels occupied by the cells there and divide it by the total area of the region. This gives us the spatial cell-density of the suspension at a given instant of time.

\section{Delaunay triangulation of the cell-positions}
\label{appendix:triangulation}

\begin{figure}
    \centering  
    \includegraphics[width=1\linewidth]{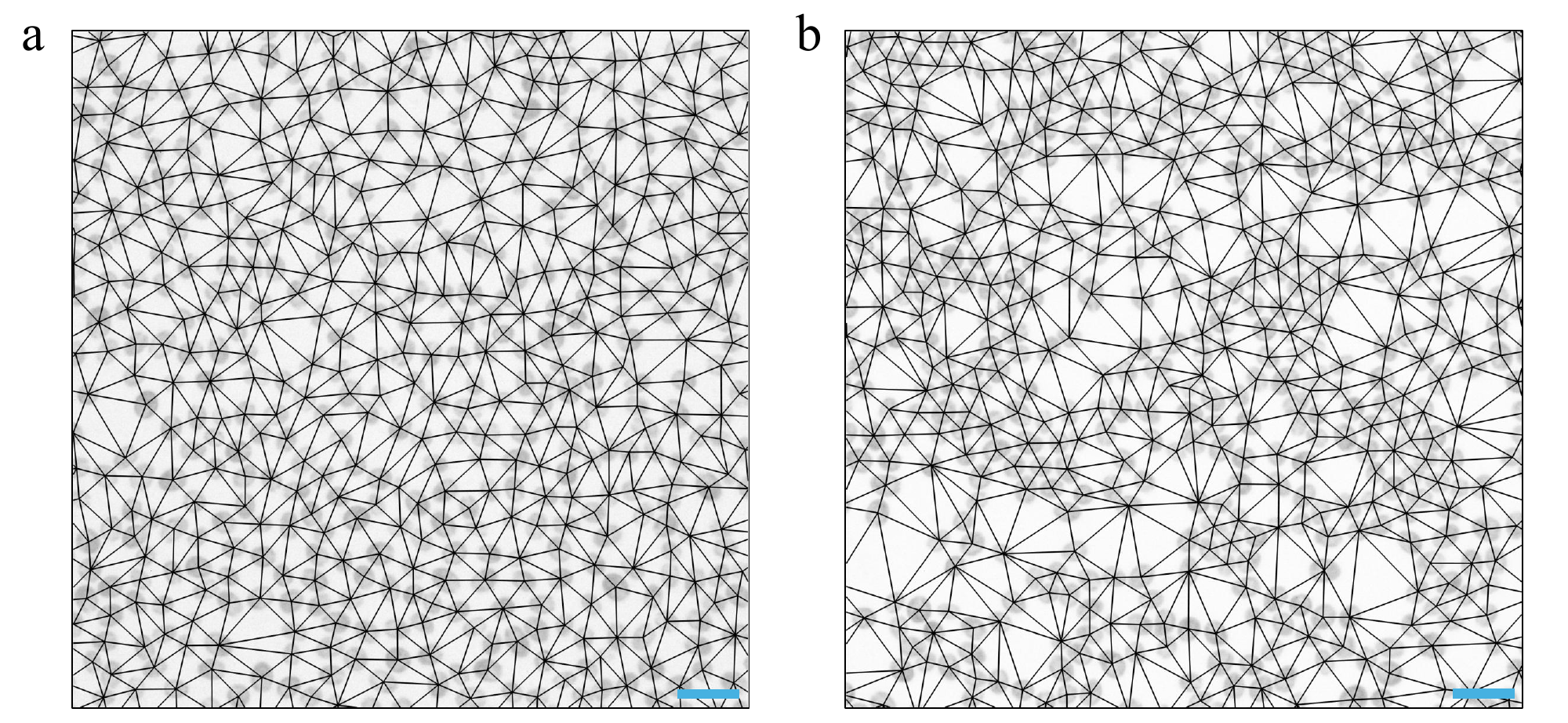}
    \caption{(a) Optical image of a WT cell suspension ($\bar{\rho} = 0.50$) overlaid with Delaunay triangles. (b) Optical image of a mbo2 cell suspension ($\bar{\rho} = 0.50$) overlaid with Delaunay triangles. (Scale-bar, 20 microns)}
    \label{fig:triangulation}
\end{figure}

Figure.~\ref{fig:triangulation} shows the Delaunay triangulation of the cell-positions for WT and mbo2 suspensions ($\bar{\rho} = 0.50$). We calculate the mean area of the triangles $\mu_T$, the median $M_T$ and the standard deviation $\varsigma_T$. For the WT cells, $\mu_T = 51.2 \mu m^2$, $M_T = 47.3\mu m^2$ and $\varsigma_T = 19.1\mu m^2$. In case of the mbo2 cells, $\mu_T = 52.0 \mu m^2$, $M_T = 41.5\mu m^2$ and $\varsigma_T = 31.9\mu m^2$. Even if the mean area for both cell types are close, there are significant differences in the median and the standard deviation. This difference in the spatial distribution of the cells can be visually seen; there appears to be much larger void spaces in the mbo2 cell suspensions compared to WT cells of similar $\bar \rho$. Given these void spaces in mbo2 cells, we do not interpolate the velocity vectors for $\bar \rho < 0.6$.  

\section{Probability distribution of the spatial density $\rho$}
\label{appendix:PDFrho}

\begin{figure}
    \centering  
    \includegraphics[width=1\linewidth]{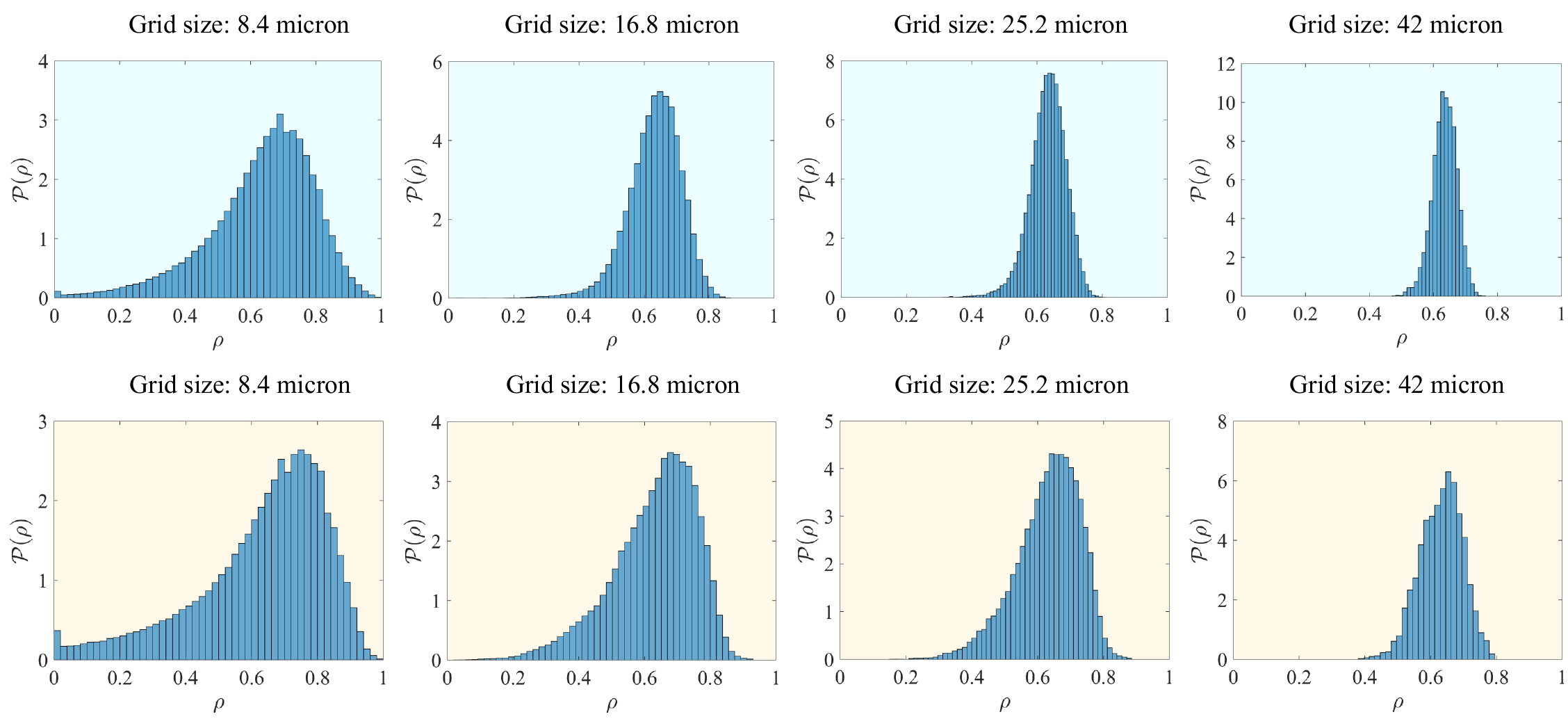}
    \caption{PDFs of the spatial density $\rho$ computed for different grid sizes of the WT cells (top panel) and mbo2 cells (bottom panel). Here, $\bar{\rho} \simeq 0.63$.}
    \label{fig:PDFrho_grid}
\end{figure}

\begin{figure}
    \centering  
    \includegraphics[width=0.8\linewidth]{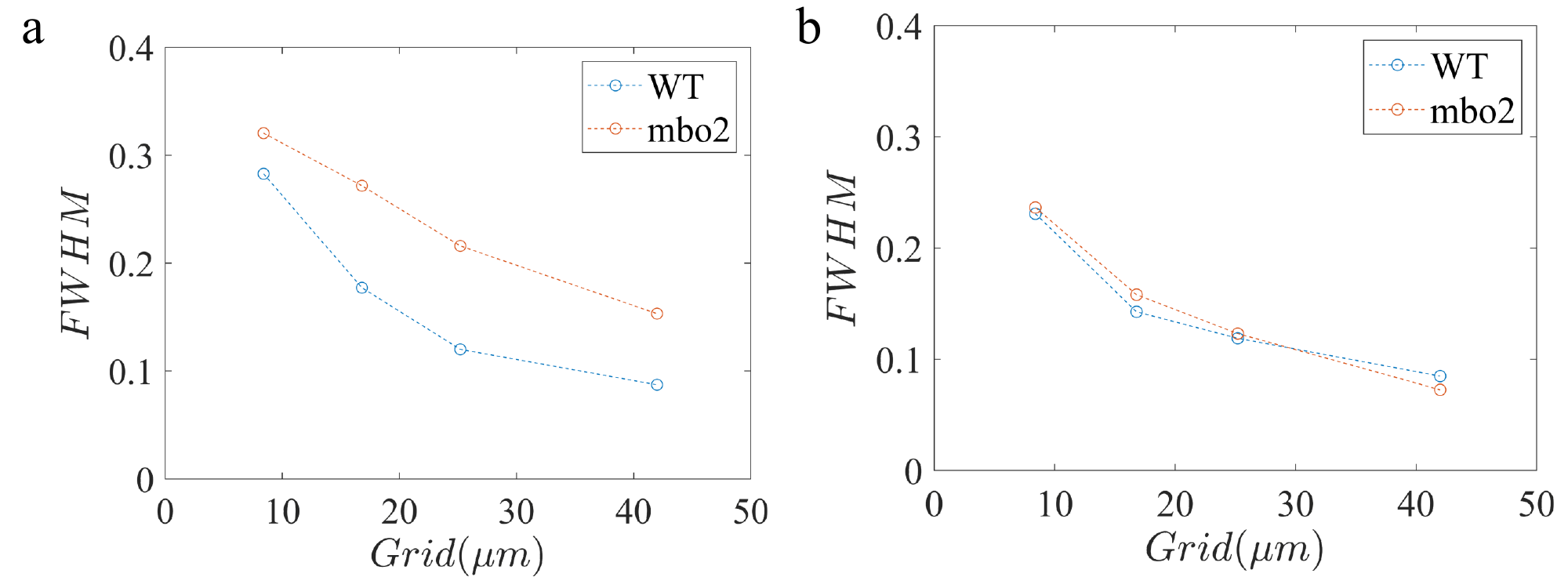}
    \caption{FWHM of the PDFs of $\rho$ for different grid sizes (a) $\bar{\rho} \simeq 0.63$ (b) $\bar{\rho} \simeq 0.75$}
    \label{fig:fwhm_grid}
\end{figure}

Figure.~\ref{fig:PDFrho_grid} plots the PDFs of the density $\rho$ for different grid sizes. They are shown for $\bar{\rho} \simeq 0.63$ for both WT and mbo2 cells. The mbo2 cells have a wider distribution compared to WT cells; and hence, they have more pronounced regions of high and low densities. But as the overall concentration increases, this difference dies down. Figure.~\ref{fig:fwhm_grid} shows the full-width at half maxima (FWHM) of the PDFs of $\rho$ versus the grid size for two different average densities. It is seen that the FWHMs are very similar at $\bar{\rho} \simeq 0.75$.


\section{Variations of velocity components with space and time}

\begin{figure}
    \centering  
    \includegraphics[width=1\linewidth]{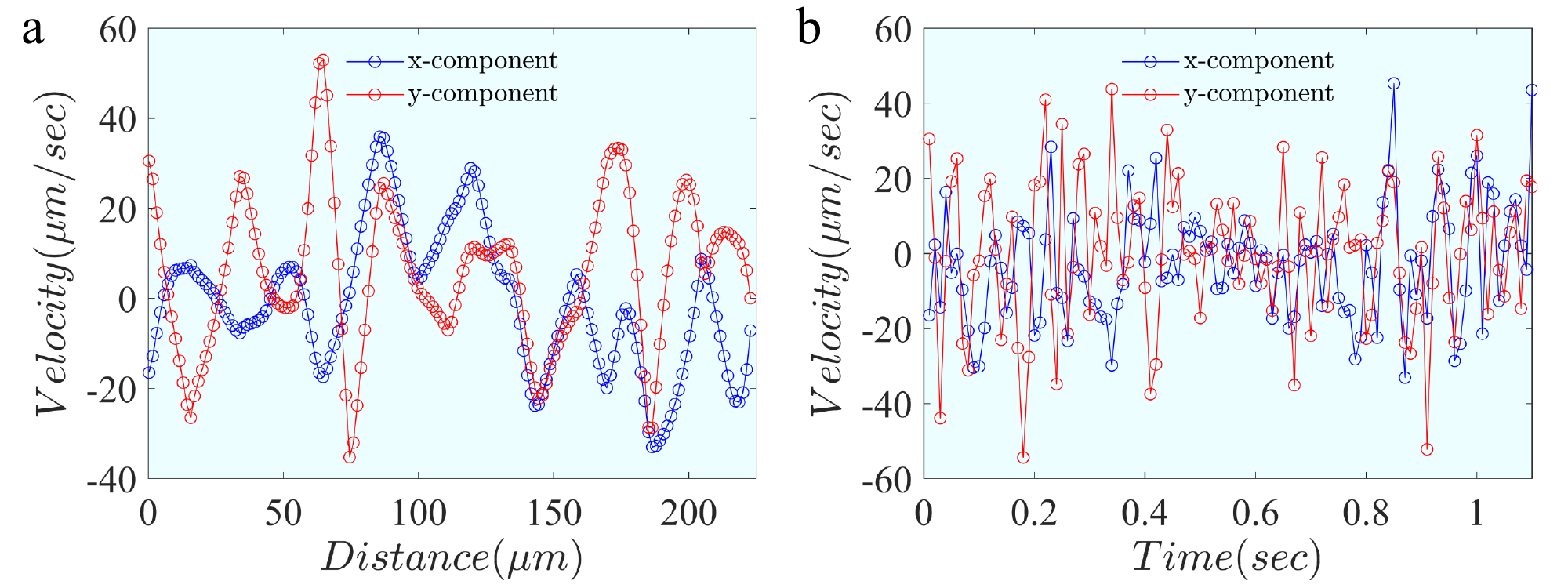}
    \caption{Plots of the x and y components of the velocity (a) versus distance, at a fixed representative time, and (b) versus time, at a fixed representative point.}
    \label{fig:velocity_space_time}
\end{figure}

Figure.~\ref{fig:velocity_space_time}(a) shows how the x and y components of velocity vary with distance. We start from a random representative point and then gradually move to the right, for a fixed frame. Figure.~\ref{fig:velocity_space_time}(b) shows how the x and y components of velocity vary with time. We start from a random representative point and plot how the velocity at that fixed point varies with time.

\bibliography{References}
\end{document}